\documentclass[aps,prd,eqsecnum,superscriptaddress,twocolumn,%
nofootinbib,floatfix,preprintnumbers]{revtex4}

\usepackage[pdftex]{graphicx}
\usepackage{amssymb}
\usepackage{amsmath}
\usepackage{longtable}
\usepackage{rotating}
\usepackage{color}
\usepackage{aeitensor}
\usepackage{bm}
\usepackage{rcs}

\RCS $Id: radiometer.tex,v 1.95 2007/12/10 19:10:32 badkri Exp $

\renewcommand{\today}{\number\year\space\ifcase\month\or
  January\or February\or March\or April\or May\or June\or
  July\or August\or September\or October\or November\or December\fi
  \space\number\day}

\newcommand{\F}{\mathcal{F}}
\newcommand{\Tobs}{T_{\text{obs}}}

\newcommand{\Tmax}{T_{\text{max}}}
\newcommand{\vect}[1]{\vec{#1}}
\newcommand{\uvec}[1]{\vec{#1}}
\newcommand{\s}{\mathcal{S}}
\newcommand{\y}{\mathcal{Y}}
\newcommand{\G}{\tilde{\mathcal{G}}}
\newcommand{\A}{\mathcal{A}}
\newcommand{\Inter}{\mathcal{I}}
\newcommand{\thr}{\rho_{\rm th}}
\newcommand{\Phase}{\Phi}

\newcommand{\nhat}{\uvec{n}}
\newcommand{\uhat}{\uvec{u}}
\newcommand{\vhat}{\uvec{v}}

\newcommand{\conj}{*}

\newcommand{\coinc}{co\"{\i}ncident}

\newcommand{\tens}[1]{\aeitensor{#1}}
\newcommand{\abs}[1]{\left\lvert#1\right\rvert}
\newcommand{\un}[1]{\text{\,#1}}
\DeclareMathOperator{\Var}{Var}
\DeclareMathOperator{\Cov}{Cov}
\DeclareMathOperator{\erfc}{erfc}

\begin{document}

\newcommand*{\AG}{Max-Planck-Institut f\"ur
    Gravitationsphysik, Albert-Einstein-Institut, Am M\"uhlenberg 1,
    D-14476 Golm, Germany}\affiliation{\AG}
\newcommand*{\IU}{Inter-University Centre for Astronomy and
  Astrophysics, Post Bag 4, Ganeshkhind, Pune 411007, India}\affiliation{\IU}

\title{The cross-correlation search for periodic gravitational waves
}

\author{Sanjeev Dhurandhar}\email{sanjeev@iucaa.ernet.in}\affiliation{\IU}
\author{Badri Krishnan}\email{badri.krishnan@aei.mpg.de}\affiliation{\AG}
\author{Himan Mukhopadhyay}\email{himan@iucaa.ernet.in}\affiliation{\IU}
\author{John T.~Whelan}\email{john.whelan@aei.mpg.de}\affiliation{\AG}

    
\begin{abstract}

  In this paper we study the use of cross-correlations between
  multiple gravitational wave (GW) data streams for detecting
  long-lived periodic signals.  Cross-correlation searches between
  data from multiple detectors have traditionally been used to search
  for stochastic GW signals, but recently they have also been used in
  directed searches for periodic GWs.  Here we further adapt the
  cross-correlation statistic for periodic GW searches by taking into
  account both the non-stationarity and the long term-phase coherence
  of the signal.  We study the statistical properties and sensitivity
  of this search, its relation to existing periodic wave searches, and
  describe the precise way in which the cross-correlation statistic
  interpolates between semi-coherent and fully-coherent methods.
  Depending on the maximum duration over we wish to preserve phase
  coherence, the cross-correlation statistic can be tuned to go from a
  standard cross-correlation statistic using data from distinct
  detectors, to the semi-coherent time-frequency methods with
  increasing coherent time baselines, and all the way to a full
  coherent search.  This leads to a unified framework for studying
  periodic wave searches and can be used to make informed trade-offs
  between computational cost, sensitivity, and robustness against
  signal uncertainties.

\end{abstract}

\preprint{LIGO-P070140-00-Z}
\preprint{AEI-2007-158}

\maketitle

\section{Introduction}

Long lived quasi-periodic gravitational waves (GWs) from rapidly
rotating non-axisymmetric neutron stars are among the promising
sources of detectable GWs for ground based detectors such as LIGO,
Virgo, GEO600 etc.  A number of searches for long-lived periodic GWs
have been carried out using data from ground based GW detectors.
These include searches using data from the interferometric and bar
detectors.  These searches are of two kinds depending on the size of
the parameter space that is searched:
\begin{description}
\item[i.] Targeted searches for sources whose parameters are
  well known from other astrophysical observations
  \cite{Abbott:2003yq,Abbott:2004ig,Abbott:2007ce}.
  Such searches are not computationally intensive, and use
  statistically optimal matched filtering techniques.
\item[ii.] Wide parameter space searches either for neutron stars in
  binary systems whose parameters are poorly constrained from prior
  observations \cite{Abbott:2006vg}, or blind searches for as yet
  unknown neutron stars
  \cite{Abbott:2006vg,Abbott:2005pu,S4PSH,S4Radiometer,Astone:2007iz}.
\end{description}
While none of the above searches have yet resulted in a detection,
there have been some notable successes.  For the searches targeting
known pulsars, the limits on the gravitational wave emission and the
corresponding limits on the deformation are starting to become
astrophysically interesting.  

Similarly, a lot of the groundwork has been laid for meeting the
computational challenges for the wide parameter space searches.
Computationally efficient methods and hierarchical data analysis
pipelines have been developed which allow us to vastly improve the
ratio of sensitivity to computational cost.  Most of these are
semi-coherent methods, i.e. combinations of coherent analyses combined
together by excess power techniques, and they come in two main
flavors.  The first combines short segments of simple Fourier
transformed data.  The baseline of the short Fourier transforms is
chosen such that the signal manifests itself as excess power in a
single frequency bin, and the excess power is combined by various
methods.  The simplest is the StackSlide method \cite{Brady:1998nj}
which adds the normalized excess power from the short segments, taking
care to ``slide'' the frequency bins to account for the Doppler shift
and intrinsic spindown.  The PowerFlux method \cite{powerflux} is very
similar; it performs a weighted sum of the normalized power using
weights which take the sky-position and polarization dependent
sensitivity of the detector into account; the weights serve to improve
the sensitivity.  Finally, there is the Hough transform method which
performs a weighted sum of binary-number counts calculated by setting
a threshold on the normalized excess power.  This is more robust and
computationally efficient, though at the cost of being somewhat less
sensitive.  All three methods have been used to analyze LIGO data in
all-sky wide frequency band searches for GWs from isolated neutron
stars\cite{Abbott:2005pu,S4PSH}, and these are so far the most
sensitive wide parameter space GW searches of their kind published so
far; we shall refer to them as the ``standard'' semi-coherent searches
in the rest of this paper.

A variant of these standard semi-coherent techniques are the so-called
hierarchical searches which aim to search deeper by increasing the
coherent time baseline
\cite{Brady:1998nj,Krishnan:2004sv,Cutler:2005pn}.  This requires a
sky-position (and spindown) dependent demodulation to be performed
before calculating the excess power statistic.  The extra demodulation
step significantly increases the computational cost and such a search pipeline
is currently being employed on larger computational platforms such as
\texttt{Einstein@Home}\footnote{http://einstein.phys.uwm.edu}.

In addition to the above surveys for isolated neutron stars, searches
have also been carried out for gravitational waves from neutron stars
in binary systems.  A plausible argument for why some neutron stars
may be emitting detectable GWs applies to neutron stars in binary
systems, and in particular, to the Low Mass X-ray Binaries (LMXBs)
which consist of a neutron star and a low mass main-sequence star.
The observed X-ray flux from these systems is due to the high rates of
accretion of matter onto the neutron star.  It is observed that the
rotation rates of neutron stars in LMXBs is significantly lower than that
might be expected on theoretical grounds; the highest theoretically
possible rotation rate is significantly larger than that of a $1\un{kHz}$,
while the current observed record is $\sim 620\un{Hz}$.  It was
suggested (first by Bildsten \cite{Bildsten:1998}) that this apparent
upper bound on the rotation rate might be due to a balance between the
spin-up due to accretion and the spindown due to the emission of
gravitational radiation - there is virtually a ``wall'' created by the
flux of GW radiated, which increases as $\Omega^6$, where $\Omega$ is
the angular rotational frequency of the spinning neutron star and this
limits its spin-up.  There are a number of other suggested
explanations which do not involve gravitational radiation, but
accreting neutron stars are clearly promising sources of detectable
gravitational radiation.  So far two searches have targeted Sco X-1,
the brightest LMXB.  These have used very different techniques;
\cite{Abbott:2006vg} used a coherent integration on 6 hours of data from
the second science run of the LIGO detectors, while
\cite{S4Radiometer} uses a cross-correlation statistic on data from
the more recent fourth science run.  The elucidation and
generalization of this cross-correlation technique tailored to
periodic GW searches, and its relation with the other searches
discussed above will occupy us for the rest of this paper.

The results from these searches are starting to become astrophysically
interesting.  For example, using data from the latest science runs of
the LIGO detectors, it is expected that the indirect spindown limit on
the amplitude of gravitational waves from the Crab pulsar will be
beaten by about a factor of 3. The resulting limits on the ellipticity
of the known pulsars are also starting to place constraints on the
equations of state of nuclear matter in neutron stars (see e.g.
\cite{Haskell:2007sh,Owen:2005fn}). A detection would lead to new
insights about neutron star physics not obtainable by other means.
Searches using large amounts of data from the LIGO detectors operating
at design sensitivity are well underway, and the results are expected
to become yet more astrophysically interesting in the near future.

Almost all of these searches mentioned above have been based on
techniques which look for signals of a given form in a single data
stream, i.e.  either matched filtering techniques or semi-coherent
power summing methods.  While both matched filtering and semi-coherent
techniques have been generalized and used to analyze data from
multiple interferometers \cite{Cutler:2005hc,S4PSH}, the starting
point for these methods is always the analysis of a single data
stream.  There is however one exception, which is the method used in
\cite{Ballmer:2005uw,S4Radiometer} and is inherently based on looking
at multiple data streams.  Let us consider two data segments
\begin{subequations}
  \label{eq:32}
  \begin{align}
    x_1(t)& \quad \textrm{for} \quad t\in [T_1-\Delta{T}/2, T_1+\Delta{T}/2]
    \ ,  \\
    x_2(t)& \quad \textrm{for} \quad t\in [T_2-\Delta{T}/2, T_2+\Delta{T}/2]
    \ .
  \end{align}  
\end{subequations}
If a signal resulting from the same gravitational wave
is present in both streams, it should be
possible to cross-correlate the output of two detectors to extract the
signal.  The basic cross-correlation statistic is
\begin{equation}
  \label{eq:31}
  \int_{T_1 - \Delta{T}/2}^{T_1+\Delta{T}/2} dt_1
  \int_{T_2-\Delta{T}/2}^{T_2+\Delta{T}/2} dt_2 \, x_1(t_1)x_2(t_2)Q(t_1,t_2) \,,
\end{equation}
where $Q(t_1,t_2)$ is an appropriately chosen filter function.  This
technique was originally developed for the stochastic background
searches where the cross-correlation is absolutely essential and is
based on the fact that multiple detectors will see the same GW signal
\cite{Allen:1997ad,Mitra:2007mc}, and it has been used extensively to
search for a stochastic GW background using LIGO data
\cite{Abbott:2003hr,Abbott:2005ez,Abbott:2006zx}.  The function
$Q(t_1,t_2)$ can be tuned to search for GWs coming from a particular
sky position and also polarization \cite{Mitra:2007mc} 
and this method has been used to search for periodic
waves from the neutron star in Sco~X-1.  All previous discussions of
this method have however been in the context of stochastic searches.
In this paper, we investigate in detail its applications for periodic
wave searches.

The optimal form of the function $Q(t_1,t_2)$ depends on the kinds of
sources that we are looking for.  Thus for a stochastic background we
use the facts that the statistical properties of the signal are time
independent and that the two polarizations are statistically
independent.  In particular, the optimal $Q$ is time invariant, i.e. a
function of only the difference $t_1-t_2$.  Furthermore, $Q$ turns out
to depend on the expected spectrum of the stochastic background.  

For periodic GWs from neutron stars, many of these assumptions do not
hold. The signal is deterministic and non-stationary (because of the
Doppler shift), and the two polarizations are not independent.
There is yet another ingredient present for periodic signals that is not
present for stochastic sources.  In principle, since the signals we
are looking for have long term phase coherence, it should be possible
to cross-correlate any pair of data segments to extract the signal,
regardless of how far apart the segments are in time and regardless of
whether they are from the same interferometer or not.  It will turn
out that the sky-resolution is much coarser than for the standard
periodic searches; the appropriate baseline is not the Earth-Sun
distance but rather the distance between the two detectors.  This
leads to a much lighter computational burden for a blind
search.  All of these issues will be discussed in detail in the rest of
this paper.  

The paper is organized as follows.  Sec.~\ref{sec:notation} sets up
notation and describes the waveforms that we are looking for; this
includes both isolated neutron stars and neutron stars in binary
systems.  It also discusses the short segment Fourier transforms
(SFTs) and the restrictions on their time baseline for the signal
power to be concentrated in a single SFT frequency bin.
Sec.~\ref{sec:formalism} motivates and defines the basic
cross-correlation statistic for a pair of short data segments;
Sec.~\ref{sec:sensitivity} discusses the statistical properties and
the sensitivity of the search; Sec.~\ref{sec:fstat} elucidates the
relation of the cross-correlation method with the $\F$ statistic;
Sec.~\ref{sec:paramest} provides estimates of the parameter estimation
that can be achieved and Sec.~\ref{sec:params} investigates the
question of resolution of parameters such as sky position, spin-down
etc.  Sec.~\ref{sec:discussion} concludes with a summary of our
results and suggestions for future work, and finally appendix
\ref{sec:generalstat} discusses some technical and conceptual issues
which have been ignored in the earlier sections for simplicity.

\section{Notation and useful equations}
\label{sec:notation}

\subsection{The waveform}
\label{subsec:waveform}

The waveform we are looking for is a tensor metric perturbation
\begin{equation}
  \tens{h}(t) = h_+(t) \tens{e}_{\!+} + h_\times(t)\tens{e}_{\!\times}
\end{equation}
where $\{\tens{e}_{\!A}|A=+,\times\}$ is a transverse-traceless
polarization basis associated with the GW propagation direction and
tailored to the polarization state of the waves so that
\begin{equation}
  \label{eq:12}
  h_+(t) = A_+\cos\Phase(t)
  \ ,
  \qquad
  h_\times(t) = A_\times\sin\Phase(t)
  \ .
\end{equation}
If $\iota$ is the angle between the line of sight $\nhat$ to the star
and its rotation axis, the amplitudes are
\begin{subequations}
\label{eq:17}  
\begin{gather}
  A_+ = h_0 \A_+
  \ ,
  \qquad
  A_\times = h_0\A_\times 
  \ ,
  \\
  \A_+ = \frac{1+\cos^2\iota}{2}
  \ ,
  \qquad
  \A_\times = \cos\iota
  \ .
\end{gather}
\end{subequations}
In the neutron star rest frame with proper time $\tau$, the phase is
\begin{equation}
  \label{eq:13}
  \Phase(t(\tau)) = \Phase_0 + 2\pi \left\{f_0\tau + \frac{1}{2}f_1\tau^2 \ldots
  \right\}
  \ .
\end{equation}
The reference time where all the spindown parameters are defined is
taken to be $\tau = 0$, and $\Phase_0$ is the phase at $\tau=0$.

A detector's scalar strain response is the contraction of the tensor
metric perturbation with a response tensor\footnote{For an
  interferometer with arms along the unit vectors $\uhat$ and $\vhat$,
  $\tens{d}=\frac{1}{2}(\uhat\otimes\uhat-\vhat\otimes\vhat)$.} $\tens{d}$:
\begin{equation}
  \label{eq:16}
  h(t) = \tens{h}(t):\tens{d}(t) = \sum_{A=+,\times}F_A(t)h_A(t)
\end{equation}
where
\begin{equation}
  F_A(t) = \tens{e}_{\!A} : \tens{d}(t)
\end{equation}
The polarization basis $\{\tens{e}_{\!A}\}$ is sometimes inconvenient,
because its definition involves not only the direction to the source
but also the source's polarization state (specifically the orientation
of the neutron star's spin).  For a given sky direction $\nhat$, one
can always construct a transverse, traceless polarization basis
$\tens{\varepsilon}_{\!A}$ by starting e.g., with the vector transverse
to $\nhat$ and lying in the Earth's equatorial plane.  The
relationship between this reference basis and the preferred
polarization basis of the source is described by the polarization
angle $\psi$:
\begin{subequations}
  \begin{align}
    \tens{e}_{\!+} &= \tens{\varepsilon}_{\!+} \cos 2\psi +
    \tens{\varepsilon}_{\!\times} \sin 2\psi \\
    \tens{e}_{\!\times} &= - \tens{\varepsilon}_{\!+} \sin 2\psi +
    \tens{\varepsilon}_{\!\times} \cos 2\psi
  \end{align}
\end{subequations}
That means, if we define
\begin{subequations}
  \label{eq:abdef}
  \begin{align}
    a(t; \nhat)&=\tens{d}(t):\tens{\varepsilon}_{\!+} (\nhat) \\
    b(t; \nhat)&=\tens{d}(t):\tens{\varepsilon}_{\!\times} (\nhat)
  \end{align}
\end{subequations}
(which are time-dependent because of the rotation of the detector
tensor $\tens{d}$), we can decompose the beam pattern functions as
\begin{subequations}
  \begin{align}
  \label{eq:29}
  F_+(t; \nhat, \psi) &= a(t; \nhat)\cos 2\psi + b(t; \nhat)\sin 2\psi
  \ ,
  \\ 
  F_\times(t; \nhat, \psi) &= b(t; \nhat)\cos 2\psi - a(t; \nhat)\sin 2\psi
  \ .
  \end{align}
\end{subequations}
The polarization angle is a property of the source, but the functions
$a(t; \nhat)$ and $b(t; \nhat)$ depend on both the sky position of the source and
the detector in question.

\subsubsection{Isolated neutron stars}
\label{subsubsec:isolated}

The relation between the detector time $t$ and the neutron star time
$\tau$ depends on whether the neutron star is isolated or in a binary.
For an isolated neutron star, we assume\footnote{As it turns out, so
long as the neutron star is moving inertially, this assumption is not
necessary; the frequencies involved are all simply offset by the
constant Doppler shift between the neutron star rest frame and the SSB.}
that the star is at rest with
respect to the SSB frame. Let $\vect{r}(t)$ be the position of the
detector in the SSB frame and $\vect{v}(t)$ its velocity.  The times
of arrival of the wave at the detector and the SSB are
\begin{equation}
  \label{eq:14}
  t = \tau - \frac{\vect{r}\cdot\nhat}{c} +
  \textrm{relativistic corrections}
  \ .
\end{equation}
The relativistic corrections can be ignored for our purposes.  The
instantaneous frequency is then, to a very good approximation
\begin{align}
  \label{eq:15}
  f(t) &= \hat{f}(t) + \hat{f}(t)\frac{\vect{v}\cdot\nhat}{c}
  \ ,
  \\
  \hat{f}(t) &= f_0 + f_1t
  \ .
\end{align}
The parameters of the signal from an isolated neutron star are thus
the so-called amplitude parameters (or nuisance parameters)
$\{h_0,\cos\iota,\psi,\Phase_0\}$ and the Doppler parameters $\bm\lambda
= \{\nhat,f_0,f_1,\ldots\}$.  The Doppler parameters determine the
frequency evolution of the signal through \eqref{eq:15}.  The
frequency and spindown ranges will canonically be taken to be
$50\un{Hz} < f_0 < 1000\un{Hz}$, and $-1\times 10^{-8}\un{Hz/s} < f_1
< 0$.  These were the ranges used in \cite{S4PSH}.  The lowest
frequency is determined by the performance of the detector, and it
will be lower for the advanced detectors.  The upper end of the
frequency range could conceivably be as high as $2000\un{Hz}$
depending on the computational cost.

Written in terms of the detector time $t$, and including first
spindowns, the phase is:
\begin{equation}
  \label{eq:20}
  \Phase(t) = \Phase_0 + 2\pi\left(f_0t + \frac{1}{2}f_1t^2\right) + 2\pi (f_0+f_1t)
  \frac{\vect{r}\cdot\nhat}{c}
  \ . 
\end{equation}
We have ignored the $\frac{1}{2}f_1(\vect{r}\cdot\nhat/c)^2$ term.  In
fact, even the term $f_1t(\vect{r}\cdot\nhat/c)$ will be ignored in
most of the calculations below.\footnote{All of these approximations are
  used only for our calculations in this paper. The actual searches do
  not make any of these approximations, and nor do they ignore the
  relativistic Einstein and Shapiro corrections.}

Let us quantify the restrictions on the parameter space due to these
approximations adapting the ``$1/4$-cycle criterion'' used in
\cite{Jaranowski:1998qm}: any physical effect which contributes less
than $1/4$ of a cycle to the phase of the signal over a given coherent
observation time will be ignored.  Since $\abs{\vect{r}\cdot\nhat/c} \leq
1\un{AU}/c \approx 500\un{s}$, we will have
$\frac{1}{2}\abs{f_1}(\vect{r}\cdot\nhat/c)^2 < 1/4$ if
$\abs{f_1} < 2\times 10^{-6}\un{Hz/s}$.  This is much larger than
any spindowns we can realistically consider.  On the other hand, the
$f_1t(\vect{r}\cdot\nhat/c)$ is, in general not negligible for
realistic spindowns and observation times of months.  However, we will
break up our observation time into shorter segments of duration much
less than a day.  Over say 1 hour, this term is ignorable if $\abs{
  f_1} < 3\times 10^{-7}\un{Hz/s}$ which is still a very large
spindown.

\subsubsection{Neutron stars in binary systems}
\label{subsubsec:binary}

To account for the motion of the neutron star in a binary orbit, we
need to add the orbital time delays to \eqref{eq:14}.  The most
important contribution is again the Roemer delay:
\begin{equation}
  \label{eq:38}
  t = \tau - \frac{\vect{r}\cdot\nhat}{c}  +
  \frac{\vect{r}_{\rm orb}\cdot\nhat}{c} +\textrm{relativistic corrections}
  \ .
\end{equation}
Here $\vect{r}_{\rm orb}$ is the position vector of the neutron star in the
binary system's center of mass frame.

There are four relevant orbital parameters.  The first is the orbital
period $P_{\rm orb}$, and, if available, its derivative $\dot{P}_{\rm
  orb}$.  We then need a reference time within the orbit for which we
use $T_{\rm asc}$, the time of crossing of the
ascending node.  The third parameter is
the projected semi-major axis of the neutron star, $a_{\rm p} = a_{\rm
  x} \sin i$.  The final parameter is the orbital eccentricity $e$.
In addition, there are 2 parameters specifying the orientation of the
orbital plane, i.e. the inclination angle $i$ (not to be confused with
the orientation of the neutron star axis $\iota$) and the argument of
periapsis $\omega$.  Of these 6 parameters, only 5 are required to
define the phase model because of the projection along the line of
sight $\nhat$; see \cite{Dhurandhar:2000sd} for further details.

We therefore have a total of 5 parameters of the binary which
determine the frequency evolution of the signal: $\bm\lambda_{\rm bin}
= (a_{\rm x}\sin i,e,P_{\rm orb},T_{\rm asc},\omega)$.  In the case
when the orbit is circular ($e=0$), the argument of periapsis and the
initial orbital phase combine additively into a single parameter so
that we are left with only 3 search parameters: $\bm\lambda_{\rm bin}
= (a_{\rm p},P_{\rm orb},T_{\rm asc})$.  We will not include higher
derivatives of ${P}_{\rm orb}$.  As an example, for Sco~X-1 (the
brightest LMXB), some of the orbital parameters are $P_{\rm orb}
\approx 6.8\times 10^4\un{s}$, and $a_{\rm p}/c \approx 1.44\un{s}$,
and $e < 3\times 10^{-3}$
\cite{Wright:1975,Steeghs:2002,Abbott:2006vg}.

Let $\vect{v}_{\rm orb}$ be the velocity of the neutron star in the
center-of-mass frame of the binary.  The observed frequency is, to a
very good approximation, given again by the non-relativistic
expression,
\begin{equation}
  \label{eq:11}
  f(t) = \hat{f}(t) + \hat{f}(t)\frac{(\vect{v} - \vect{v}_{\rm
      orb})\cdot\nhat}{c}
  \ . 
\end{equation}
Since $\vect{v}_{\rm orb}$ is usually much larger than the Earth's
orbital velocity, $\vect{v}_{\rm orb}$ is the dominant contribution to
the Doppler shift.  

\subsection{Short-time Fourier transforms}
\label{subsec:sfts}

Given a time series detector output from a detector, it is convenient
to break it up into short segments of length $\Delta{T}$ and to store
the Short-time Fourier Transforms (SFTs).  The value of $\Delta{T}$ is
chosen such that the approximation \eqref{eq:19} is valid and as we
will see, this leads to different restrictions on $\Delta{T}$ for
neutron stars which are isolated or in binary systems.  Such SFT
databases are commonly used in the LIGO, GEO and Virgo collaborations
for periodic wave searches, and we will also base our data analysis
strategies mostly on SFTs \cite{v2SFTs}.

Let $x(t)$ be a time series sampled discretely at intervals of $\delta
t$.  Let us consider $N$ samples $x_j$ for $j=0\ldots N-1$, and let
$\Delta{T} = N\delta t$.  Our convention for the discrete Fourier
transform will be
\begin{equation}
  \label{eq:35}
  \tilde{x}_k = \delta t \sum_{j=0}^{N-1}x_je^{-i2\pi jk/N}
  \ ,
\end{equation}
where $k=0,1\ldots (N-1)$.  For $0\leq k \leq \lfloor N/2 \rfloor$,
the frequency index $k$ corresponds to a physical frequency $f_k=
k/\Delta{T}$ with $\lfloor .\rfloor$ denoting the integer part of a
given real number.  The values $\lfloor N/2 \rfloor < k \leq N-1$
correspond to negative frequencies given by $f_k = (k-N)/\Delta{T}$.
Each SFT stores the real and imaginary values of $\tilde{x}_k$ for a
range of frequency bin indices $k$. The $I^{\text{th}}$ SFT will span the
time interval $[T_I - \Delta{T}/2, T_I + \Delta{T}/2]$.  When necessary,
we will denote the data at the $k^{\text{th}}$
frequency bin of the $I^{\text{th}}$ SFT by $\tilde{x}_{k,I}$.  

Eq.\eqref{eq:35} is actually a simplification.  In practice, to
avoid spectral leakage, a taper $w_j$ is applied while taking the
Fourier transform: 
\begin{equation}
  \label{eq:40}
  \tilde{x}_k = \sum_{j=0}^{N-1} w_j x_j e^{-i2\pi jk/N}
  \ .
\end{equation}
See e.g. \cite{Percival-Walden} for details. We will mostly ignore
window-related issues in this paper.  

The detector output $x(t)$ is the sum of noise $n(t)$ plus a possible
gravitational wave signal:
\begin{equation}
  \label{eq:36}
  x(t) = n(t) + h(t)
  \ .                 
\end{equation}
We will assume the noise to be a real stochastic process of zero mean,
stationary and Gaussian; in practice, we only need stationarity over a
period $\Delta{T}$, the time baseline of the SFTs. The properties of
the noise are thus fully described by a single-sided power
spectral density $S_n(f)$ which, in the continuous time case is
defined as,
\begin{equation}
  \label{eq:37}
  S_n(f) := 2 \int_{-\infty}^{\infty} \langle n(t'+t)n(t')\rangle
  e^{-i2\pi ft} dt
  \ ,
\end{equation}
where $\langle \cdot \rangle$ denotes an average over an ensemble of
noise realizations.  Note that the average $\langle
n(t'+t)n(t') \rangle$ is independent of $t'$ because
of the assumption of stationarity.  In practice, we are of course only
given $x(t)$ and not $n(t)$ itself.  So we must take care to ensure
that the estimation of $S_n(f)$ is not biased by the presence of a
signal.  Finally, the following expression for $S_n$ is useful:
\begin{equation}
  \label{eq:1}
  \langle |\tilde{x}_k|^2\rangle \approx \frac{\Delta{T}}{2}S_n(f_k)
  \ .
\end{equation}
This equation relates the variance of the (real and imaginary parts)
of $\tilde{n}_k$ to the PSD, thus providing a more intuitive
understanding of the PSD. 
This is a special case of a more general expression which, in the
continuous case, reads,
\begin{equation}
  \label{eq:2}
  \langle \tilde{n}^\conj(f) \tilde{n}(f')\rangle =
  \frac{1}{2}S_n(f) \delta(f-f')
  \ .
\end{equation}

\subsection{The short-duration Fourier transform of the signal}
\label{subsec:fourier}

We now calculate the Fourier transform of the signal over an
observation duration $[T-\Delta{T}/2,T+\Delta{T}/2]$ centered at the
time $T$.  We assume $\Delta{T}$ is small enough so that $\{F_A|A=+,\times\}$
can be treated as constants in this duration; this means $\Delta{T} \ll
1\un{day}$.  We assume that the observation duration is small enough so
that the phase of the signal in this duration can be expanded in a
power series at the mid-point $T$:
\begin{equation}
  \label{eq:18}
  \Phase(t) = \Phase(T) + 2\pi f(T)(t-T)
  \ .
\end{equation}
The validity of this approximation sets the limits on how large
$\Delta{T}$ can be.  If $\dot{f}(t)$ is the time-derivative of the
signal frequency at any given time $t$, the above approximation is
valid whenever effects of the frequency derivative $\dot{f}$ can be
ignored over the duration $\Delta{T}$. Using the $1/4$-cycle criterion,
this leads to $\dot{f} \leq \Delta{T}^{-2}$.

For isolated neutron stars, the time variation of $f(t)$ is given by
\eqref{eq:15} and is due to two effects: the intrinsic spindown
of the star, and the Doppler modulation due to the Earth's motion.
Consider first the intrinsic spindown $f_1$.  Taking the largest
spindown to be $10^{-8}\un{Hz/s}$, we get $\Delta{T} < 10^4\un{s}$.  For the
Doppler shift, we can estimate $\dot{f}$ by keeping $\hat{f}$ fixed
and differentiating $\vect{v}$ in \eqref{eq:15}.  The result is
worked out in \cite{Krishnan:2004sv} and yields the following
restriction on $\Delta{T}$:
\begin{equation}
  \label{eq:34}
  \Delta{T} < 4\times 10^3 \un{s} \times
  \sqrt{\frac{500\un{Hz}}{f_0}}
  \ .  
\end{equation}
In this paper, for isolated neutron stars, we will mostly use
$\Delta{T}=30\un{min}$ as a canonical reference value. This is well within the
above restrictions.  The limits on $\Delta{T}$ are far more stringent
for neutron stars in binary systems because of the higher Doppler
shifts.  The Sco~X-1 search in \cite{Abbott:2006vg} used $\Delta{T} =
60\un{s}$.

With the approximation \eqref{eq:18}, in the time interval
$[T-\Delta{T}/2,T+\Delta{T}/2]$ we have,
\begin{equation}
  \label{eq:19}
  \begin{split}
  h(t) =& F_+A_+\cos(\Phase(T) + 2\pi f(T)(t-T)) \\
  &+ F_\times A_\times\sin(\Phase(T) + 2\pi f(T)(t-T))
  \ . 
  \end{split}
\end{equation}
The Fourier transform of $h(t)$ is easily seen to be,
\begin{multline}
  \label{eq:21}
  \tilde{h}(f) = \int_{T-\Delta{T}/2}^{T+\Delta{T}/2}
  h(t)e^{-i2\pi  f(t-T+\Delta{T}/2)} dt \\
  = e^{i\pi f\Delta{T}}
  \Biggl[
  e^{i\Phase(T)}\frac{(F_+A_+ -iF_\times A_\times)}{2}
  \delta_{\Delta{T}}\left( f-f(T)\right)  \\ 
  + e^{-i\Phase(T)}\frac{(F_+A_+ +iF_\times A_\times)}{2}
  \delta_{\Delta{T}}\left(f+f(T)\right)    
  \Biggr]
  \,,
\end{multline}
where we have defined the finite time approximation
$\delta_{\Delta{T}}(f) := \sin(\pi f\Delta{T})/\pi f$ to the delta function
$\delta(f)$.  This definition of the function $\delta_{\Delta{T}}(f)$
leads to significant spectral leakage of the signal power into
neighboring frequency bins. This can be improved by using suitable
tapers as in \eqref{eq:40}. We assume that this has been done and
we will henceforth assume that spectral leakage is negligible.

\section{The cross-correlation statistic for a pair of SFTs}
\label{sec:formalism}

Let us assume that we have two data streams covering the time
intervals $\Inter_I$ and $\Inter_J$ centered on the times $T_I$ and
$T_J$ respectively; both intervals have the same duration $\Delta{T}$.
The data streams in the two intervals $x_I$ and $x_J$ could come from
the same or different detectors, though of course if $T_I=T_J$ then
the detectors have to be different.  The received signals in the two
intervals are denoted by $h_I(t)$ ($t\in\Inter_I$) and $h_J(t)$
($t\in\Inter_J$) respectively.  As before, we assume that the duration
$\Delta{T}$ of the time intervals is such that the beam pattern
functions are approximately constant.  We denote the PSDs of the noise
in the two intervals by $S_n^{(I)}(f)$ and $S_n^{(J)}(f)$
respectively.

The basic cross-correlation statistic corresponding to a filter
function $Q$ is,
\begin{equation}
  \label{eq:4}
  \s_{I\!J} = \int_{T_I-\Delta{T}/2}^{T_I+\Delta{T}/2}dt
  \int_{T_J-\Delta{T}/2}^{T_J+\Delta{T}/2}dt'\,
  x_I(t)x_J(t')Q_{I\!J}(t,t')
  \ .
\end{equation}
We would like to understand how the optimal $Q_{I\!J}$ can be chosen.
The optimal choice depends in fact on the kind of signals we are
looking for.  The analysis presented in \cite{Allen:1997ad} describes
the optimal choice of $Q$ for stochastic signals, and here we will
tailor our discussion to the periodic signals described earlier.

To get some intuition on the nature of $\s_{I\!J}$, let us evaluate
$\s_{I\!J}$ in the frequency domain assuming that $Q_{I\!J}$ is time
invariant: $Q(t,t') = Q(t-t')$. Keep in mind however that
this will \emph{not} be the optimal solution, and a more detailed
analysis will be presented later.

It is easy to evaluate \eqref{eq:4} by writing $x_I(t)$ in terms
of its Fourier transform.  Along the way we approximate
$\delta_{\Delta{T}}$ by the delta function, but we however should not
take $Q_{I\!J}(\tau)$ to be a rapidly decreasing function of $\tau$ as
in \cite{Allen:1997ad}.  Since our signals have long term phase
coherence, $Q_{I\!J}(\tau)$ will also turn out to be periodic.  In any
case, we still end up with the simple expression,
\begin{equation}
  \label{eq:9}
  \s_{I\!J} = \int_{-\infty}^{\infty}df\,
  \tilde{x}_I^\conj(f)\tilde{x}_J(f)\tilde{Q}_{I\!J}(f)
  \ .
\end{equation}
The mean value of $\s_{I\!J}$ over an ensemble of noise realizations is,
\begin{equation}
  \label{eq:10}
  \mu_{I\!J} := \langle\s_{I\!J}\rangle =
  \int_{-\infty}^{\infty}df
  \,\tilde{h}_I^\conj(f)\tilde{h}_J(f)\tilde{Q}_{I\!J}(f)
  \ . 
\end{equation}
Here we have assumed that the noise has zero mean, and that $n_I$ and
$n_J$ are uncorrelated.  If we assume further that $h_I \ll n_I$ then
the standard deviation is approximately:
\begin{equation}
  \label{eq:23}
  \sigma_{I\!J}^2 = \frac{\Delta{T}}{2}\int_0^\infty
  df\,S_{n}^{(I)}(f)S_{n}^{(J)}(f)|\tilde{Q}_{I\!J}(f)|^2
  \ .
\end{equation}
Furthermore, it can also be shown under the same assumptions, that
$\s_{I\!J}$ and $\s_{J\!K}$ are uncorrelated for $K\neq I$:
\begin{equation}
  \label{eq:3}
  \langle \s_{I\!J}\s_{J\!K}\rangle = \delta_{I\!K}\sigma^2_{I\!J}
  \ .
\end{equation}
Thus, the correlation pairs formed from all pairs of distinct SFTs are
statistically independent.  Note however that the same is not true for
the third order moments; for example $\langle \s_{I\!J}\s_{J\!K}\s_{K\!I}
\rangle \neq 0$ even when the small signal approximation is valid.
This is however not a problem for us because we will never need to
calculate the third and higher order correlations between the $\{\s_{I\!J}\}$.



\par
Eq.\eqref{eq:10} clearly demonstrates that taking
$Q_{I\!J}(t,t')$ to be time-invariant is, in general, suboptimal
for the data analysis problem at hand. The signal frequencies $f_I =
f(T_I)$ and $f_J = f(T_J)$ at the midpoints of the two intervals are
given by \eqref{eq:15} (for an isolated system) or \eqref{eq:11}
(for a binary systems).  In general, $f_I$ and $f_J$ may be quite
distinct from each other, especially if the intervals are far apart in
time.  Our assumptions on $\Delta{T}$ ensure that the signal power to
be concentrated mostly in a single SFT frequency bin.  Thus, no matter
what we choose for $\tilde{Q}_{I\!J}(f)$, the overlap between
$\tilde{h}_I$ and $\tilde{h}_J$ might be quite small.  This will lead
to a small $\mu_{I\!J}$ and thus a small signal-to-noise ratio
$\mu_{I\!J}/\sigma_{I\!J}$.  The fix is obvious: we need to shift the
frequencies while constructing the cross-correlation statistic.  So,
if we define $\delta f_{I\!J} = f_J - f_I$ then,
\begin{equation}
  \label{eq:5}
  \s_{I\!J} =
  \int_{-\infty}^{\infty}df\,\tilde{x}_I^\conj(f)
  \tilde{x}_J(f + \delta f_{I\!J})\tilde{Q}_{I\!J}(f + \delta f_{I\!J}/2)
  \ .
\end{equation}
In the time domain, this corresponds to the non-time invariant filter:
\begin{equation}
  \label{eq:6}
  Q_{I\!J}(t,t') = e^{-i\pi (\delta f_{I\!J}) (t+t')}Q_{I\!J}(t-t')
  \ .
\end{equation}
The mean $\mu_{I\!J}$ becomes,
\begin{equation}
  \label{eq:7}
  \mu_{I\!J} := \langle\s_{I\!J}\rangle =
  \int_{-\infty}^{\infty}df\,\tilde{h}_I^\conj(f)
  \tilde{h}_J(f + \delta f_{I\!J})\tilde{Q}_{I\!J}(f + \delta f_{I\!J}/2)
  \ ,
\end{equation}
and the variance $\sigma_{I\!J}^2$ is unchanged.

An important quantity for us is the signal cross-correlation
$\tilde{h}^\conj_I(f)\tilde{h}_J(f+\delta f_{I\!J})$.  We extract the
amplitude term $h_0^2$ and the delta-functions to define for $f > 0$, 
\begin{equation}
  \label{eq:8}
  \tilde{h}^\conj_I(f)\tilde{h}_J(f+\delta f_{I\!J}) = h_0^2\G_{I\!J}\delta^2_{\Delta{T}}(f-f_I)
  \ .
\end{equation}
The signal cross-correlation function $\G_{I\!J}$ is an important
quantity, much like the overlap-reduction function for stochastic
searches defined in \cite{Allen:1997ad} (though $\G_{I\!J}$ is not
exactly analogous to the overlap reduction function).

Apart from the frequency $f$ and $T_I,T_J$, $\G_{I\!J}$ is a function of
the signal parameters, i.e. the amplitude parameters
$\{h_0,\iota,\psi,\Phase_0\}$, the Doppler parameters $\bm\lambda$, and
possibly the binary parameters $\bm\lambda_{\rm bin}$.  To avoid
clutter, we will often drop the dependence of $\G_{I\!J}$ on the signal
parameters and $T_{I}$, $T_{J}$, and just write $\G_{I\!J}$. 

Using \eqref{eq:21} it is easy to calculate $\G_{I\!J}$.  For
$ f>0 $, the dominant contribution is:
\begin{widetext}
  \begin{gather}
    \label{eq:25}
    \G_{I\!J} =
    \frac{1}{4}e^{-i\Delta\Phase_{I\!J}}\left\{(F_{I+}F_{J+}\A_+^2 + 
      F_{I\times}F_{J\times}\A_\times^2) - i(F_{I+}F_{J\times} -
      F_{I\times}F_{J+})\A_+\A_\times\right\}
    \ ,
    \\
    \label{eq:22}
    \Delta\Phase_{I\!J} =  \Phase_I(T_I) - \Phase_J(T_J)
    \ .
  \end{gather}
  Here we have added the subscript $I$ and $J$ to the phase $\Phase$
  to emphasize that $\Phase$ is detector dependent.  For isolated
  neutron stars, with the approximations explained in
  Sec.~\ref{subsubsec:isolated}, this leads to
  \begin{equation}
    \label{eq:39}
    \Delta\Phase_{I\!J} =  2\pi \sum_{k=0}^s\frac{f_k}{k!} (T_I^{k+1} - T_J^{k+1}) + 2\pi f_0
    \frac{\Delta\vect{r}_{I\!J}\cdot\nhat}{c}
    \ .
  \end{equation}
\end{widetext}
We have used \eqref{eq:20}, ignored the
$f_1t(\vect{r}\cdot\nhat/c)$ term, and defined $\Delta\vect{r}_{I\!J}:=
\vect{r}(T_I)- \vect{r}(T_J)$.  Recall from \eqref{eq:17} that
$\A_{+,\times}$ are the same as $A_{+,\times}$ but without the factor
of $h_0$.

We can now also average over $\cos\iota$ using the following
relations:
\begin{subequations}
  \label{eq:26}
  \begin{gather}
    \langle \A_+^2\rangle_{\cos\iota} = \frac{7}{15}
    \ ,
    \qquad
    \langle \A_\times^2\rangle_{\cos\iota}  = \frac{1}{3}
    \ ,
    \\ 
    \langle \A_+\A_\times\rangle_{\cos\iota}= 0
    \ .
  \end{gather}
\end{subequations}
The average of $\G_{I\!J}$ over $\cos\iota$ is thus,
\begin{equation}
  \langle\G_{I\!J}\rangle_{\cos\iota} =
  \frac{1}{60}e^{- i\Delta\Phase_{I\!J}}(7F_{I+}F_{J+} +
    5F_{I\times}F_{J\times})
    \ .
    \label{eq:27}
\end{equation}
We can easily perform another average over the polarization angle
$\psi$ using \eqref{eq:29}:
\begin{equation}
  \label{eq:28}
  \begin{split}
    \langle F_{I+}F_{J+} \rangle_\psi &=   \langle F_{I\times}F_{J\times}
    \rangle_\psi = \frac{1}{2} (a_I a_J + b_I b_J)
    \\
    &= d_{Iab}\, P^{\text{TT}\nhat}{}^{ab}_{cd}\, d_J^{cd}
    \ ,
  \end{split}
\end{equation}
where,
\begin{equation}
  P^{\text{TT}\nhat}{}^{ab}_{cd}
  =\frac{1}{2}\sum_{A=+,\times}\varepsilon_{A\,ab}\varepsilon_{A}^{cd}
  =\frac{1}{2}\sum_{A=+,\times}e_{A\,ab}e_{A}^{cd}
  \ ,
\end{equation}
is a projection onto symmetric traceless tensors transverse to $\nhat$.
This leads to:
\begin{equation}
\begin{split}
  \langle\G_{I\!J}\rangle_{\cos\iota,\psi} &=
  \frac{1}{10}e^{-i\Delta\Phase_{I\!J}}(a_Ia_J + b_Ib_J)
  \ .
  \label{eq:30}
  \\
  &=
  \frac{1}{5}\,d_{Iab}\, d_J^{cd}\, P^{\text{TT}\nhat}{}^{ab}_{cd}
  \, e^{-i\Delta\Phase_{I\!J}}
  \ .
\end{split}
\end{equation}
In the case of time-{\coinc} SFTs, since $\Delta\Phase_{I\!J}$ reduces
$\frac{\Delta\vect{r}_{I\!J}\cdot\nhat}{c}$, this is just a
normalization factor times the overlap reduction function which would
be used for a search for a stochastic background coming from a single
point on the
sky.\cite{S4Radiometer,Ballmer:2005uw,JTWanisotropic,Whelan:2005sk}

\section{Statistics and sensitivity}
\label{sec:sensitivity}

For each SFT pair (labeled by an index pair $I\!J$,
we define the raw cross-correlation as the
complex random variable:
\begin{equation}
  \label{eq:42}
  \y_{k,I\!J} = \frac{\tilde{x}_{k,I}^\conj\tilde{x}_{k',J}}{\Delta{T}^2}
  \ . 
\end{equation}
The frequency bin $k'$ is shifted from $k$ by an amount corresponding
to $\delta f_{I\!J}$: $k' = k + \lfloor \Delta{T}\delta
f_{I\!J}\rfloor$.  Note that $\y_{k,I\!J}$ is computed using only data
from single frequency bins in the two SFTs; this works under the
assumption that the signal power is mostly concentrated in a single
frequency bin.  We emphasize that, this is not a fundamental limitation
because we could, if we wished, consider the (optimally weighted)
power from the neighboring bins as well if necessary.  In the rest of
this paper, we shall consider $\Delta T$ sufficiently small so that
this assumption is valid.  See Sec.~\ref{sec:notation} for
quantitative estimates on $\Delta T$.  

In this section we initially make two additional simplifying
assumptions.  First we take the signal to be much smaller than the
noise, i.e. $h \ll n$, and second we only consider $\y_{k,I\!J}$ for
$I\neq J$.\footnote{Both of these assumptions will be relaxed in
Appendix~\ref{sec:generalstat}.} The results obtained using these assumptions
are probably the most relevant for practical applications.
Firstly, for the ground based detectors the signal is indeed
expected to be much smaller than the noise.  Secondly, the number of
pairs of distinct SFTs is much more than the number of self pairs;
there is thus no significant loss in sensitivity if the
self-correlations are not considered in the final detection statistic.

The $\{\y_{k,I\!J}\}$ are random variables with mean and variance given by,
\begin{gather}
  \label{eq:24}
  \mu_{k,I\!J} = h_0^2 \G_{I\!J}
  \ ,
  \\
  \sigma_{k,I\!J}^2 = \sigma_{k,I\!J}^2
  = \frac{1}{4\Delta{T}^2}S_n^{(I)}(f_k)S_n^{(J)}(f_{k'})
  \ .
\end{gather}
To derive the expression for the mean, we have replaced
$\delta_{\Delta T}(f-f_I)$ by $\delta_{\Delta{T}}(0) = \Delta{T}$,
and for the variance we have assumed that the real and imaginary parts
of $\tilde{x}_k$ are uncorrelated and have the same variance.  The
$\{\y_{k,I\!J}\}$ are not Gaussian variables, but we will only need their
mean and standard deviation.

Where convenient, we will replace the pair $I\!J$ with a single lowercase
Greek index $\alpha,\beta\ldots$.  Thus, $\y_{k,I\!J}$ will often be
denoted $\y_{k,\alpha}$.  To avoid unnecessary clutter, we also avoid
putting the frequency index $k$ explicitly in $\y_{k,\alpha}$. In any
case, one expects the signal contribution to be limited essentially to
a single frequency bin $k$. Our task is now to combine the
$\y_{\alpha}$ in a statistically optimal way to extract the signal
amplitude $h_0$.  The following analysis is very similar to what is
used in \cite{S4PSH} (see also \cite{powerflux,wt-hough}).

We consider detection statistics which are weighted sums of the
$\y_\alpha$:
\begin{equation}
  \label{eq:41}
  \rho = \sum_{\alpha} (u_\alpha \y_\alpha + u_\alpha^\conj
  \y_\alpha^\conj)
  \ .
\end{equation}
We are interested in the probability distribution of the random 
variable $\rho$ because this is required for computing the sensitivity  
at given false alarm and false dismissal rates. It is simply obtained by 
examining the behavior of the noise in $\y_{\alpha}$ (of which $\rho$ is made up of) 
which is derived from
\eqref{eq:42} by replacing the data $x$ by the noise $n$ in each data 
segment $I, J$. If we assume that the noise in each detector is Gaussian 
with mean zero, the noise in $\rho$ is a sum of products of 
real independent Gaussian variables each having mean zero. Although $\y_{\alpha}$ is 
complex, the statistic $\rho$ is real. The product of two 
independent Gaussian variables whose mean is zero, is a random variable 
whose probability density function (PDF) is essentially $K_0(x)$, where $K_0 (x)$ 
is the modified Bessel function of the second kind of order zero - more specifically, 
if $X \sim N(0, \sigma_X)$ and $Y \sim N(0, \sigma_Y)$, then the PDF of $Z=XY$ is 
$K_0(|z|/\sigma_X \sigma_Y) / \pi \sigma_X \sigma_Y$. This distribution has zero mean 
and a finite variance, namely, $\sigma_X^2 \sigma_Y^2$. Then a generalization of the 
central limit theorem states that the sum of a large number of such zero mean variables 
tends to a Gaussian random variable \cite{Feller:1965}. Thus $\rho$ is a Gaussian random 
variable whose mean $\mu$ and variance $\sigma^2$ are given by:   
%
\begin{gather}
  \label{eq:44}
  \mu = \sum_\alpha (u_\alpha \mu_\alpha + u_\alpha^\conj
  \mu_\alpha^\conj) = h_0^2\sum_\alpha
  (u_\alpha\G_\alpha + u_\alpha^\conj\G_\alpha^\conj)
  \ ,
  \\
  \sigma^2 = 2\, \sum_\alpha |u_\alpha|^2\sigma_\alpha^2
  \ .
\end{gather}
Let us set a threshold $\thr$ on $\rho$ to select detection
candidates based on a false alarm rate $\alpha$.  It is easy to show
that for Gaussian noise the threshold must be:
\begin{equation}
  \label{eq:45}
  \thr = \sqrt{2}\sigma\erfc^{-1}(2\alpha)
  \ ,
\end{equation}
where $\erfc$ is the complementary error function.  
The detection rate in the presence of a signal is,


\begin{equation}
  \label{eq:46}
  \gamma = \frac{1}{2}\erfc\left(\frac{\thr - \mu}{\sqrt{2}\sigma}\right)
  \ .
\end{equation}
Since $\mu \propto h_0^2$, this can be inverted to give the smallest
value of $h_0$ that will cross the threshold at given false alarm
and detection rates,
\begin{equation}
  \label{eq:47}
  h_0^2 = 2\, \mathcal{S}
  \left(
    \frac{\sqrt{\sum_\alpha |u_\alpha|^2 \sigma_\alpha^2}}
    {\sum_\alpha (u_\alpha\G_\alpha + u_\alpha^\conj\G_\alpha^\conj)}
  \right)
  \ ,
\end{equation}
where $\mathcal{S} = \erfc^{-1}(2 \alpha) - \erfc^{-1}(2 \gamma)$.
This can also be written in terms of the false dismissal rate $\beta =
1-\gamma$ as $\mathcal{S} = \erfc^{-1}(2 \alpha) + \erfc^{-1}(2
\beta)$.\footnote{This is proved by using the following property of
  the complementary error function: $\erfc(-x)= 2-\erfc(x)$. Setting
  $x = -\erfc^{-1}(2\gamma)$, we get $2-2\beta=2\gamma = \erfc(-x) =
  2-\erfc(x)$, which yields $x=\erfc^{-1}(2\beta)$. } The solution for
$u_\alpha$ which minimizes $h_0$ can then be shown to be\footnote{This
  is perhaps easiest to see if we define a positive-definite
  inner-product over vectors $\mathbf{x} = \{x_\alpha\}$ as
  $\mathbf{x}\cdot\mathbf{y} := \sum_{\alpha}\textrm{Re}\left[
    x_\alpha^\conj y_\alpha\right] \sigma_\alpha^2$. In terms of this
  inner product (\ref{eq:47}) can be written as $h_0 =
  \mathcal{S}\frac{||\mathbf{u}||}{\mathbf{u}\cdot\mathbf{H}}$ where
  $H_\alpha = \G_\alpha^\conj/\sigma_\alpha^2$. $h_0$ is then minimum when
  $\mathbf{u}$ is parallel to $\mathbf{H}$.},
\begin{equation}
  \label{eq:48}
  u_\alpha \propto \frac{\G_\alpha^\conj}{\sigma_\alpha^2}
  \ .
\end{equation}
It is shown in appendix \ref{sec:generalstat} that this solution also
holds when we include the self-correlations (still assuming $h\ll n$).

Substituting from (\ref{eq:48}) into (\ref{eq:41}), the optimal
detection statistic is: 
\begin{equation}
  \label{eq:54}
  \rho  \propto
  \sum_\alpha \frac{\y_\alpha\G_\alpha^\conj + \y_\alpha^\conj\G_\alpha}
  {\sigma_\alpha^2}\ .
\end{equation}
Substituting the expression for $u_\alpha$ from \eqref{eq:48} back
into \eqref{eq:47}, the optimal sensitivity is seen to be,
\begin{equation}
  \label{eq:49}
  h_0 = \left( \frac{\mathcal{S}^2}{\sum_\alpha
      |\G_\alpha|^2/\sigma_\alpha^2}\right)^{1/4}
  \ . 
\end{equation}
In the case when we are correlating data from two distinct
interferometers with stationary noise floors $S_n^{(1)}(f)$ and
$S_{n}^{(2)}(f)$, then $\sigma_\alpha$ is independent of $\alpha$ and
is given by,
\begin{equation}
  \label{eq:50}
  \sigma_\alpha^2 = \frac{1}{4\Delta{T}^2}S_n^{(1)}(f)S_n^{(2)}(f)
  \ .
\end{equation}
We are using the superscripts in $S_n^{(1)}$ and $S_N^{(2)}$ to refer
to the two detectors.  Similarly, if we denote the average of
$|\G_\alpha|^2$ over pairs of SFTs by
$\langle|\G_\alpha|^2\rangle_\alpha$, then,
\begin{equation}
  \label{eq:55}
  \sum_{\alpha} |\G_\alpha|^2 = N_{\rm pairs} \langle
  |\G_\alpha|^2\rangle_\alpha
\end{equation}
where $N_{\rm pairs}$ is the total number of SFT pairs.  This leads to,
\begin{equation}
  \label{eq:51}
  h_0 = \frac{\mathcal{S}^{1/2}}{\sqrt{2}\langle
    |\G_\alpha|^2\rangle_\alpha^{1/4}} \frac{1}{N_{\rm pairs}^{1/4}}
  \sqrt{\frac{\left(S_n^{(1)}S_n^{(2)}\right)^{1/2}}{\Delta{T}}}
  \ . 
\end{equation}
Similarly, if we relax the requirement that the pairs have to be from
the distinct detectors, and instead assume that the noise floor in all
SFTs is the same, $S_n$, then 
\begin{equation}
  \label{eq:92}
   h_0 = \frac{\mathcal{S}^{1/2}}{\sqrt{2}\langle
    |\G_\alpha|^2\rangle_\alpha^{1/4}} \frac{1}{N_{\rm pairs}^{1/4}}
  \sqrt{\frac{S_n}{\Delta{T}}}
  \ . 
\end{equation}
These are the equation we were after. They give us the sensitivity of
the cross-correlation search as a function of the statistical false
alarm and false dismissal rates, the SFT baseline $\Delta{T}$, the
noise floors of the SFTs, the number of SFT pairs $N_{\rm pairs}$, and
the geometrical factors contained in $\G_\alpha$.  They tells us that
the sensitivity grows coherently with $\Delta{T}$ and incoherently
with $N_{\rm pairs}$.  Note however that we can correlate any SFT pair
we like, so that $N_{\rm pairs}$ can be made much larger than the
number of SFTs $N_{\rm sft}$ (even if we were to exclude
self-correlations).  In fact, if we believe the signal to maintain
phase coherence over the entire observation time (which may be months
or years), and if we can afford to do so computationally, then $N_{\rm
  pairs} \sim N_{\rm sft}^2$ so that $h_0 \propto (N_{\rm
  sft}\Delta{T})^{-1/2}$ which is better than what we would get with
the standard semi-coherent searches \cite{S4PSH}.

\section{The relation with the $\F$ statistic}
\label{sec:fstat}

From \eqref{eq:51}, we see that if we use all SFT pairs available,
the amplitude sensitivity of the cross-correlation search is proportional to
$\Tobs^{-1/2}$ which is what we would get for a fully coherent
search.  There must thus be a close relation between the
cross-correlation and the coherent matched filter, and in this section
we show that this is indeed the case.

A convenient implementation of the matched filter statistic for
periodic waves is provided by the so-called $\F$-statistic first
defined in \cite{Jaranowski:1998qm} for the single interferometer
case, and later generalized to the multi-interferometer case in
\cite{Cutler:2005hc}, and a detailed study of the parameter space
resolution was presented in \cite{Prix:2006wm}.  Let us start with the
single interferometer case.  

For defining the $\F$-statistic, it is convenient to rewrite the
waveform of \eqref{eq:12}. We first separate out the initial phase
$\Phase_0$ from the total phase as, 
\begin{equation}
  \label{eq:60}
  \Phase(t) = \Phase_0 + \varphi(t)
  \,.
\end{equation}
We decompose the total waveform $h(t)$ in terms of four quadratures
as,
\begin{equation}
  \label{eq:61}
  h(t) =\sum_{i=1}^{4} \A^\mu h_\mu(t)
  \ ,
\end{equation}
where the four amplitudes $\{\A^\mu\}$ (not to be confused with $\A_+$ and
$\A_\times$) are time independent and the $\{h_\mu\}$ are
\begin{equation}
  \begin{split}
    \label{eq:63}
    h_1(t) = a(t)\cos\varphi(t)
    \ ,
    &\qquad
    h_2(t) = b(t)\cos\varphi(t)
    \,,
    \\
    h_3(t) = a(t)\sin\varphi(t)
    \ ,
    &\qquad
    h_4(t) = b(t)\sin\varphi(t)
    \ ,
  \end{split}
\end{equation}
with $a(t)$ and $b(t)$ defined as in \eqref{eq:abdef}.
What this decomposition achieves is a separation of the amplitude
parameters $\{h_0,\iota,\psi,\Phase_0\}$ from the Doppler parameters.  The
only signal parameters in the quadratures $\{h_\mu\}$ are the Doppler
parameters while the amplitudes $\{\A^\mu\}$ depend only on the amplitude
parameters.

In order to extract the signal $h(t)$ from the noise, the optimal
search statistic is the likelihood function $\Lambda$ defined by, 
\begin{equation}
  \label{eq:64}
  \ln \Lambda = (x|h)-\frac{1}{2}(h|h)
  \ ,
\end{equation}
where the inner product $(\cdot |\cdot)$ is defined as:
\begin{equation}
  \label{eq:65}
  (x|y) := 2 \int_0^\infty \frac{\tilde{x}(f)\tilde{y}^\conj(f) +
    \tilde{x}^\conj(f)\tilde{y}(f)}{S_n(f)} \, df
  \ .
\end{equation}
The quantity $\ln\Lambda$ is essentially the matched filter and is
precisely what we should use in order to best detect the waveform
$h(t)$.  An explicit search over the amplitude parameters $\{\A^\mu\}$ is
avoided by noting that $\ln\Lambda$ depends quadratically on the
$\{\A^\mu\}$.  We can thus analytically find the maximum likelihood (ML)
estimators $\{\widehat{\A}^\mu\}$ of the amplitudes $\{\A^\mu\}$
by solving the set of four coupled linear equations:
\begin{equation}
  \label{eq:66}
  \left.
    \frac{\partial \ln\Lambda}{\partial \A^\mu}
  \right\rvert_{\A^\nu=\widehat{\A}^\nu} = 0
  \ ,
  \qquad
  \mu=1,\ldots,4
  \ .
\end{equation}
The $\F$-statistic is then defined as the log likelihood ratio with the
values of the amplitudes $\{\A^\mu\}$ replaced by their ML estimators:
\begin{equation}
  \label{eq:67}
  \F := \left.\ln\Lambda\right\rvert_{\A^\mu=\widehat{\A}^\mu}
  \ .
\end{equation}
Explicitly, $\F$ can be written as
\begin{equation}
  \label{eq:68}
  \F = \frac{4}{S_n(f_0)} \frac{B|F_a|^2 + A|F_b|^2 -
    C(F_aF_b^\conj + F_bF_a^\conj)}{AB - C^2}
  \ ,
\end{equation}
where 
\begin{subequations}
  \begin{align}
    \label{eq:62}
    F_a &= \int_{0}^{\Tobs} x(t)a(t)e^{-i\varphi(t)}dt
    \ ,
    \\
    F_b &= \int_{0}^{\Tobs} x(t)b(t)e^{-i\varphi(t)}dt
    \ ,
    \\
    A &= \int_0^{\Tobs} a^2(t)\,dt
    \ ,
    \quad
    B = \int_0^{\Tobs}b^2(t)\,dt
    \ ,
    \\
    C &= \int_0^{\Tobs} a(t)b(t)\,dt
    \ .
  \end{align}
\end{subequations}
We need to write the $F_a$ and $F_b$ still more explicitly; let us
start with $F_a$. We break up the integral for $F_a$ into sub-intervals
defined by the SFTs, and assume as we have been doing all along that
$a(t)$ is constant over the SFT duration:
\begin{multline}
  \label{eq:72}
  F_a = \sum_{I} \int_{T_I - \Delta{T}/2}^{T_I +
    \Delta{T}/2} x(t)a(t)e^{-i\varphi(t)}dt  \\
  = \sum_{I} a_I\int_{T_I - \Delta{T}/2}^{T_I +
    \Delta{T}/2} x(t)e^{-i\varphi(t)}dt
  \ .
\end{multline}
Writing the phase in a Taylor series around the SFT mid-time and
keeping the linear terms, we get,
\begin{equation}
  \label{eq:69}
  \varphi(t) = \varphi(T_I) + i 2\pi f_I (t-T_I)
  \ ,
\end{equation}
which leads to,
\begin{multline}
  \label{eq:70}
  F_a = \sum_{I} a_Ie^{-i\varphi(T_I)}
  \int_{T_I - \Delta{T}/2}^{T_I + \Delta{T}/2}
  x(t)e^{-i2\pi f_I (t-T_I)}dt \\
  = \sum_{I} a_Ie^{-i\varphi(T_I)}
  e^{-i\pi f\Delta{T}}\tilde{x}_I(f_I)
  \ ,
\end{multline}
and likewise for $F_b$.

Now we are ready to look at $\F$ again.  From \eqref{eq:68} it is
clear that $\F$ is quadratic in the data and from \eqref{eq:70} it
is clear that we will end up with an expansion like,
\begin{equation}
  \label{eq:71}
  \F = \sum_{I\!J} \F_{I\!J}
  \ .
\end{equation}
In fact, it turns out that \eqref{eq:71} is precisely a linear
combination of the $\y_\alpha$ defined in \eqref{eq:42}. Explicitly, it 
follows from \eqref{eq:70} that:
\begin{equation}
  \label{eq:73}
  |F_a|^2
  = \sum_{I\!J} a_I a_J
  \left(
    e^{i \Delta\Phase_{I\!J}} \y_{I\!J}
    + e^{-i \Delta\Phase_{I\!J}} \y_{I\!J}^\conj
  \right)
  \ . 
\end{equation}
Similar expressions are obtained for $|F_b|^2$ and the cross term
$F_aF_b^\conj + F_bF_a^\conj$ of the $\F$ statistic. Combining all of
the results from above, we see that $\F$ is a detection statistic of
the form \eqref{eq:41} with weights,
\begin{equation}
  u_{I\!J} \propto \left(Ab_Ib_J + Ba_Ia_J - C(a_Ib_J + a_Jb_I)\right)
  e^{i\Delta\Phase_{I\!J}}
  \ . 
\end{equation}
In the case where $A\approx B$ and $C\ll A,B$, this is seen to be
proportional to $\G_{I\!J}$ averaged over $\cos\iota$ and $\psi$
\eqref{eq:30}.  Thus we see that the cross-correlation statistic
$\rho$ is indeed roughly equivalent to the $\mathcal{F}$-statistic.
In principle, $\rho$ using the full signal cross-correlation function
$\G_\alpha$ from \eqref{eq:25}, is a function of the Doppler
parameters and also of $\{A_+,A_\times,\psi\}$; this is more like the
likelihood-ratio (modulo the dependence on the initial phase $\Phase_0$)
before maximizing it over the amplitude parameters to obtain the
$\F$-statistic.  The $\rho$ calculated with
$\langle\G_\alpha\rangle_{\cos\iota,\psi}$ is closer to the matched
filter statistic marginalized over $\cos\iota$ and $\psi$.

\section{Estimating the amplitude parameters}
\label{sec:paramest}

Thus far, we have focused on constructing the cross-correlation
statistic which is optimal for the detecting the presence of periodic
GWs.  Thus, the choice of weights given in
\eqref{eq:48} is tailored for measurements of excess
cross-correlation power, and is not actually an estimator for the
signal amplitude. Estimating the Doppler parameters
$\{f_0,f_1,\ldots,\nhat\}$ is easy since we are searching over these
parameters explicitly.  Note also that the signal cross-correlation
function $\G_{\alpha}$ of \eqref{eq:25} is a function of
$\cos\iota$ and $\psi$. We could thus, in principle, find the values
of $\cos\iota$ and $\psi$ which maximize $\rho$, thus yielding
estimators of these quantities. In practice however, we expect it to
be more convenient to use a single statistic, such as that associated
with the averaged $\G_\alpha$ given in
\eqref{eq:30}, and then estimate $\{A_+, A_\times,\psi\}$
in a follow-up stage.\footnote{Note that the
  cross-correlations $\y_\alpha$ are independent of the initial phase
  $\Phase_0$.  Thus, it is not possible to estimate $\Phase_0$ if we
  restrict ourselves to measurements of $\y_\alpha$.}  In this
section, we show that it is indeed possible to estimate
$\{A_+,A_\times,\psi\}$.  The method presented here is a straightforward
generalization of \cite{Mendell:2007ww} (see also
\cite{powerflux,S4PSH}) developed for the standard semi-coherent
searches.

The basic idea is to note that the two polarizations $h_+$ and
$h_\times$ appear in the detector with different amplitude
modulations.  Therefore, given sufficient measurements of the
$\y_\alpha$, it should be possible to extract the signal components
with different amplitude modulation patterns thereby estimating the
amplitudes $A_+$ and $A_\times$.  Let us start by defining the signal
cross-correlation functions $\G_\alpha^+$ and $\G_\alpha^\times$ for
the two polarizations which are analogous to $\G_\alpha$:
\begin{subequations}
  \begin{align}
    \label{eq:75}
    \G_{I\!J}^+ &= \frac{1}{4}e^{-i\Delta\Phase_{I\!J}}F_{I+}F_{J+}
    \ ,
    \\
    \G_{I\!J}^\times &= \frac{1}{4}e^{-i\Delta\Phase_{I\!J}}F_{I\times}F_{J\times}
    \ .
  \end{align}
\end{subequations}
These functions are significant because, just as in \eqref{eq:24},
they tell us about the mean $\mu_\alpha$ of $\y_\alpha$ for the two
independent polarizations.  The contributions of $A_+$ and $A_\times$
to the mean are respectively:
\begin{equation}
  \label{eq:79}
  \mu_\alpha^+ = A_+^2\G_\alpha^+
  \quad \textrm{and} \quad
  \mu_\alpha^\times =
  A_\times^2\G_\alpha^\times
  \ . 
\end{equation}
An estimator of $A_+$ is obtained by minimizing the following
$\chi^2$-statistic as a function of $A_+^2$,
\begin{equation}
  \label{eq:76}
  \chi^2 = \sum_\alpha \frac{\abs{\y_\alpha -
      A_+^2\G_\alpha^+}^2}{\sigma_\alpha^2}
  \ .
\end{equation}
The solution of $\partial\chi^2/\partial A_+^2 = 0$ is easily seen to be, 
\begin{equation}
  \label{eq:77}
  A_+^2 = \left(\sum_\beta
    \frac{2|\G_\beta^+|^2}{\sigma_\beta^2}\right)^{-1} 
  \sum_\alpha \frac{\y_\alpha^\conj\G_\alpha^+ +
      \y_\alpha\G_\alpha^{+\conj}}{\sigma_\alpha^2}
    \ .
\end{equation}
Similarly, the estimator for $A_\times$ is, 
\begin{equation}
  \label{eq:78}
    A_\times^2 = \left(\sum_\beta
    \frac{2|\G_\beta^\times|^2}{\sigma_\beta^2}\right)^{-1} 
  \sum_\alpha \frac{\y_\alpha^\conj\G_\alpha^\times +
      \y_\alpha\G_\alpha^{\times\conj}}{\sigma_\alpha^2}
    \ .
\end{equation}
Since $\{\G_\alpha^A\}$ depend on the polarization angle $\psi$
through the beam pattern functions, both \eqref{eq:77} and
\eqref{eq:78} imply a search over $\psi$.  We expect these estimators
to be better than the ones used in the standard semi-coherent methods
simply because it uses a larger number of measurements including all
possible pairs of SFTs.  Note that these estimators for $A_+^2$ and
$A_\times^2$ are proportional to the optimal excess-power statistic
$\rho$ of \eqref{eq:54}, with $\G_\alpha$ replaced by
$\G_\alpha^+$ and $\G_\alpha^\times$.

Finally, while we do not discuss it here, following
\cite{Mendell:2007ww}, this discussion can be generalized to construct
a joint $\chi^2$ statistic for $A_+^2$, $A_\times^2$ and $\psi$ for a
general elliptically polarized signal.

\section{Parameter space resolution}
\label{sec:params}

In this section we discuss the parameter space resolution required for
the cross-correlation statistic $\rho$.  This affects the
astrophysical significance of the search in terms of parameter
estimation and also the computational requirements for carrying out
the search.  The parameter space resolution for a detection statistic
$\rho$ is usually discussed in terms of the parameter space metric.
This is defined as the fractional loss in the signal-to-noise ratio
when $\rho$ is calculated at a point in parameter space which is
slightly different from the point corresponding to the actual signal
parameters \cite{Sathyaprakash:1991mt,Dhurandhar:1992mw,Owen:1995tm}.
In our case, we are in principle free to consider any subset of all
the possible SFT pairs in calculating the final detection statistic
$\rho$.  However, without some control on which SFT pairs are chosen,
it seems very hard to get a handle on the parameter space metric for
the general cross-correlation statistic $\rho$ defined by
(\ref{eq:41}). Our suggestion is the following: Choose a time duration
$T_{\rm max}$ and include only those SFT pairs $\{I,J\}$ for which
$|T_I - T_J| \leq T_{\rm max}$.  Thus, $T_{\rm max}$ can be viewed as
the maximum duration over which we choose to maintain strict phase
coherence.

If $T_{\rm max} = T_{\rm obs}$, then we are including all possible
pairs, and at the other extreme, if $T_{\rm max}= 0$ then we are
including only self-correlations and time-coincident correlations
between different detectors.  In the intermediate regime the
cross-correlation search is closest in spirit to a semi-coherent
hierarchical scheme which consists of breaking up the total data
available (say from $t=0$ to $t= T_{\rm obs}$ into shorter segments
$[0,T_{\rm max}]\,,[T_{\rm max},2T_{\rm max}]\ldots$.  One then
performs a coherent analysis on each of the segments (using, say, the
$\F$-statistic) and combines the results semi-coherently
\cite{Brady:1998nj,Krishnan:2004sv,Cutler:2005pn}.  The pair selection
criteria would lead us to choose all possible SFT pairs within each of
the segments.  Since we have already seen in Sec.~\ref{sec:fstat} that
this is essentially equivalent to the $\F$-statistic, the similarities
between the two schemes is obvious.  The two schemes are however not
exactly identical because this SFT pair selection criteria also
includes choosing pairs lying in adjacent data segments (assuming the
segments are sufficiently close to each other).  Thus, the
cross-correlation search with coherence time $T_{\rm max}$ will be
more sensitive than the semi-coherent search with coherent segments of
duration $T_{\rm max}$ but the precise improvement depends on the duty
cycle of the detectors, i.e. on the gaps between the SFTs and the
coherent segments.

With this criteria of choosing pairs, we will see that the resolution
depends on $T_{\rm max}$ the SFT baseline $\Delta{T}$. To make our
results concrete, we will focus on the ground based interferometers by
taking the frequency range to be from $50\un{Hz}$ to $1000\un{Hz}$.
Given the similarities with the semi-coherent and hierarchical schemes
discussed above, it is clear that a proper discussion of the metric
requires a calculation of the parameter space metric for semi-coherent
searches.  This is a combination of the coherent metric worked out in
detail in \cite{Brady:1997ji,Prix:2006wm}, and the semi-coherent
metric obtained by summing $\F$-statistic segments.  Preliminary
calculations have been worked out in \cite{Brady:1998nj}, but a
detailed study of its properties is still lacking.  We will instead
resort to order of magnitude estimates (which, in spite of their
approximate nature, have actually turned out to be fairly useful for
previous searches; see e.g. \cite{Krishnan:2004sv}).

We can either use the amplitude modulation of the detection statistic
$\rho = \sum_\alpha\rho_\alpha$ by which we mean the variation of
$\rho_\alpha$ with $\alpha$, or we can use the frequency modulation
reflected in the different frequency bins $k$ and $k^\prime$ used to
calculate the cross correlation $\y_\alpha =
\tilde{x}_{k,I}^\conj\tilde{x}_{k^\prime,J}$.  Starting with the
sky-resolution, we identify three factors which could be relevant: the
detector beam pattern functions, the detector-pair baseline
$\Delta\vect{r}_{I\!J}$, and the Doppler information over a duration
$\Delta T$ and $T_{\rm max}$; we discuss all of these in turn. The
relative importance of these three factors depends on the search
parameters.
\begin{description}
\item[i.] The expectation value of the cross-correlation statistic
  varies with the SFT pair index $\alpha$, and part of this variation
  is due to the geometrical factor $a_Ia_J + b_Ib_J$ in \eqref{eq:22}.
  Since this variation depends on the sky-position, it can in
  principle be used to get sky-position information.  The resolution
  thus obtained is roughly given by the angular scales over which the
  beam pattern functions vary.  Note that this amplitude modulation is
  due to the rotation of Earth around its axis; this is independent of
  the signal frequency and gets mostly averaged out if $\Delta{T}$ is
  comparable or larger than a day.

\item[ii.] The other reason for the variation of the SNR with $\alpha$ is
  the $\Delta \vect{r}_{I\!J}$ term in \eqref{eq:22}.  In the case
  when the two SFTs are {\coinc} in time ($T_I = T_J$), then
  $\Delta\vect{r}_{I\!J}$ is the separation between the two detectors;
  for the LIGO Hanford and Livingston observatories, this corresponds
  to a light travel time of about $10\un{ms}$.  More generally, the
  magnitude of $\Delta \vect{r}_{I\!J}$ is the distance between the
  positions of the two (distinct or same) detectors at different
  times; it could be as much as $2\un{AU}$ if $T_I - T_J \sim
  6\un{months}$.  On the other extreme, it could be zero if we are
  correlating the data with itself (which is what the standard
  semi-coherent methods do); this effect then becomes completely
  irrelevant.  If $\lambda_{\rm gw}$ is the wavelength of the wave we
  are trying to detect, the sky-resolution associated with
  $\Delta\vect{r}_{I\!J}$ is inversely proportional to the frequency:
  \begin{equation}
    \label{eq:56}
    (\Delta \theta)_{\Delta \vect{r}} \approx \frac{\lambda_{\rm
          gw}}{|\Delta\vect{r}|} = \frac{1}{f\cdot|\Delta\vect{r}|/c}
      \ . 
  \end{equation}  
  For the Hanford-Livingston pair, this corresponds to about
  $\mathcal{O}(60^{\circ})$ at $100\un{Hz}$ and about $6^\circ$ at
  $1000\un{Hz}$.

\item[iii.] The third way of getting sky-position information is
  through the Doppler shift.  This is only useful if the frequency
  resolution of the SFTs is small enough; the maximum Doppler shift is
  $f |\vect{v}|/c $, so for the Doppler shift to be important, we must
  have,
  \begin{equation}
    \label{eq:57}
    \Delta{T} > \frac{\lambda_{\rm gw}}{|\vect{v}|}
    \ .
  \end{equation}
  The magnitude of Earth's orbital velocity in its orbit is $\sim
  10^{-4}c$, so \eqref{eq:57} leads to $\Delta{T} > 200\un{s}$ at
  $50\un{Hz}$ and $\Delta{T} > 6.67\un{s}$ at $1500\un{Hz}$.  One
  relevant baseline in this case is the distance traveled by the
  detector in the duration $\Delta{T}$. Thus, the sky resolution is
  (see \cite{Krishnan:2004sv} for further details):
  \begin{equation}
    \label{eq:58}
    (\Delta\theta)_{\rm doppler} = \frac{\lambda_{\rm gw}}{|\vect{v}|
      \Delta{T}}
    \ .
  \end{equation}
  For $1800\un{s}$ SFTs, this corresponds to $\sim 6^\circ$ at
  $50\un{Hz}$ and $0.2^\circ$ at $1500\un{Hz}$. There is finally the
  baseline corresponding to $T_{\rm max}$, i.e. the distance $d_{\rm
    max}$ traveled by the detector during $T_{\rm max}$.  This leads
  to 
  \begin{equation}
    \label{eq:90}
    (\Delta\theta)_{\rm doppler} = \frac{\lambda_{\rm gw}}{d_{\rm max}}\ .
  \end{equation}
  More generally, the resolution corresponding to $T_{\rm max}$ (for
  sufficiently large $\Delta T$) is precisely the coherent metric
  calculated in \cite{Brady:1997ji,Prix:2006wm}.
\end{description}
We see that the first two items above can be viewed as using the
amplitude modulation information (dependence of the SNR on the pair
index $\alpha$), while the third uses the frequency modulation.

Let us now discuss the resolution in spindown parameters $f_k$. The
spindown term in $\Delta\Phase_{I\!J}$ appears in the combination
$f_k(T_I^{k+1} - T_J^{k+1})$. Thus, it is clear that for $T_I\neq T_J$
this leads to a spindown resolution of,
\begin{equation}
  \label{eq:59}
  (\delta f_k)_{\rm min} = \frac{1}{{\rm
      max}_{I,J}\left\{\abs{T_I^{k+1} - T_J^{k+1}}\right\}} 
\end{equation}
Thus, if we were to consider all possible pairs from a given set of
SFTs, and if we define the reference time to be in the mid-point of
the observation duration, then we would have $\delta f_k \propto
\Tobs^{-(k+1)}$.  

We can also consider the frequency resolution $(\delta f)_{\rm sft} =
(\Delta T)^{-1}$ of the SFTs themselves.  The corresponding
resolution in $f_k$ is defined by the smallest change in $f_k$
required to change the frequency by a $(\delta f)_{\rm sft}$ over the
full observation time $T_{\rm obs}$.  This leads to $\delta f_k =
(\delta f)_{\rm sft}/T_{\rm obs}^k$ for $k=1,2\ldots$.

Let us conclude this section by giving a short numerical example for
the case when we correlate data from a pair of spatially separated
detectors at the same times.  We consider frequencies of $100\un{Hz}$
and $1000\un{Hz}$, and two sky positions: one at the celestial equator
and one at $45^\circ$ degrees above it. In each case we consider
sources with the optimal orientation $\iota=0$, without any spindown
parameters, and with $\psi = 0$. The total observation time is taken
to be $\Tobs = 1\un{yr}$ and the SFT baseline is $\Delta T =
30\un{min}$.  We assume that the two data streams are coming from the
LIGO Livingston and Hanford interferometers.  For performing the
cross-correlations, we use,
\begin{equation}
  Q (t;f,\nhat) = \lambda(t;\nhat) \langle \G(\nhat) \rangle_{\cos \iota, \psi},
\label{eq:beam}
\end{equation}
where, $\lambda(t;\nhat)$ is a proportionality constant. We consider
essentially identical time segments - same barycentric time - in the
two detectors.  In a year's worth of observation time there are little
over 17,000 such time segments, each of 30 minutes duration. Thus the
time-segment indices $I, J$ each, sequentially run over the full
observation time. The relevant quantities $Q, \G$ and $\lambda$ in
\eqref{eq:beam} which carry the same indices also do the same over
the observation time - thus we may think of each of them as functions
of $t$ - the segment time-stamp; thus $I$ or $J$ is replaced by $t$.

\begin{figure*}
  \centering
  \includegraphics[width=\textwidth]{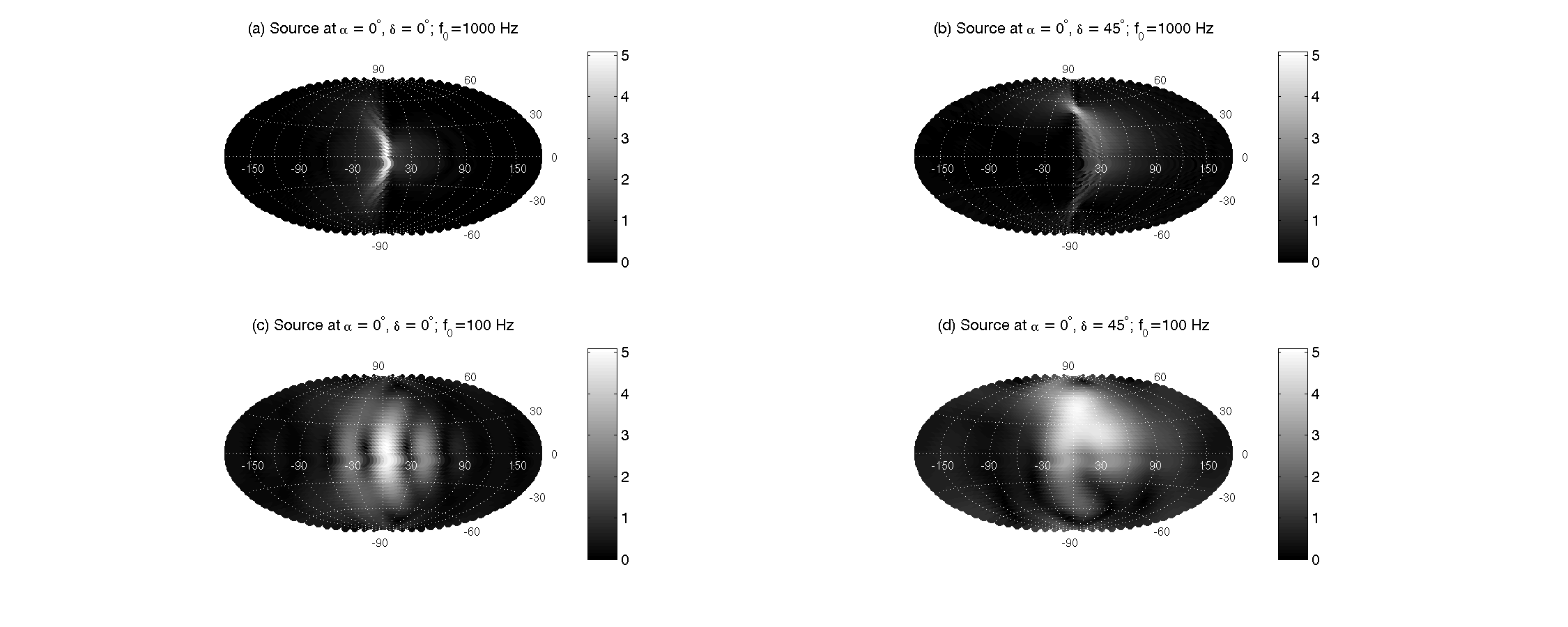}
  \caption{The point spread functions (PSFs) for sources with
    frequencies $100\un{Hz}$ [(c) and (d)] and $1000\un{Hz}$ [(a) and (b)].
    The source is taken at the celestial equator [(a) and (c)]
    and $45^\circ$ above the celestial equator [(b) and (d)]. In
    all the cases, the source orientation is taken to be optimal,
    i.e., $\iota = \psi = 0$.}
\label{fig:beam}
\end{figure*}

For the signal only case, the cross-correlation can be written
explicitly as:
\begin{equation}
  B(\nhat,\nhat') = \Lambda(\nhat)  \int_0^{T_obs} dt\,
  \langle\G(t; \nhat)\rangle_{\cos \iota, \psi} h_{(1)}(t;\nhat')h_{(2)}(t;\nhat') \,,
\end{equation}
where the subscripts in $h_{(1)}$ and $h_{(2)}$ refer to the two
distinct detectors we are considering. We have chosen,
\begin{equation}
  \Lambda^{-1}(\nhat) = \frac{1}{\Delta T} \int \lambda^{-1}(t;\nhat) \, dt
  \ .
\end{equation}
We choose the proportionality constant $\lambda$ such that
it is inversely proportional to square of the average total power accessible to
the network for a particular direction of the sky in the interval $\Delta T$
of the SFTs. Thus we have,
\begin{equation}
    \lambda^{-1}(t;\nhat) = \Delta T \langle\G(t; \nhat)\rangle_{\cos \iota, \psi}^2.
\end{equation}
This is in the spirit of the normalization scheme adopted in \cite{Mitra:2007mc}.
Figure \ref{fig:beam} shows $B(\nhat, \nhat')$ evaluated numerically
for point sources at different positions. We note that the maximum value of $B$ is 5. This 
is the result of the average value of $\G$ we have chosen in defining the 
filter function together with the fact that we have chosen optimally oriented
sources for the numerical computation. The sky-resolution is characterized
quantitatively by the FWHM (full width at half maximum) of the PSF.
From the figures it turns out to be $\simeq 8^\circ$ for $f_0 =
1000\un{Hz}$ and $\simeq 80^\circ$ for $f_0 = 100\un{Hz}$. We observe
that the agreement between the order of magnitude estimates obtained earlier and the actual
values computed from the figure is satisfactory.

\section{Discussion}
\label{sec:discussion}

We summarize the main results of this paper.  We have generalized
the cross-correlation statistic, traditionally used for the stochastic
gravitational wave background searches, to periodic gravitational
waves.  The features of periodic waves, not present in the stochastic
background signals, are non-stationarity and long-term coherence.  The
non-stationarity may need to be taken into account depending on the
frequency resolution, and the long-term coherence implies that we can
in principle cross-correlate data segments from arbitrary times and
arbitrary detectors.  This makes the method very flexible, and these are
some of the possibilities:
\begin{description}
\item[i.] We can, if we wish, correlate all possible short data segments.
  If this is done, then we showed that the resulting detection
  statistic is very close to the $\F$-statistic corresponding to a
  full matched filter statistic.  This is ideally the most sensitive
  method, but it's computational cost becomes prohibitive for wide
  parameter space searches.
\item[ii.] At the other extreme, we can choose to correlate only data
  segments taken from distinct detectors at the same (or very close)
  times.  This is the closest in spirit to the standard directed
  stochastic background searches using aperture synthesis.  In this
  mode of operation, the search is not computationally intensive, and
  is very robust against signal uncertainties.  However, this also
  implies poor resolution in parameter space, and thus more expensive
  follow-ups to verify possible detections and to estimate the signal
  parameters.
\item[iii.] From the perspective of this paper, the standard semi-coherent
  searches such as PowerFlux, StackSlide and Hough all correspond to
  the special case in which we consider only self-correlations.  The
  procedure of considering weighted sums of the cross-correlation
  power is closest to the PowerFlux method.  In fact, many of the
  lessons learnt in the PowerFlux searches should be applicable here
  with suitable modifications.  For example, the estimation of the
  signal amplitudes developed originally for PowerFlux carries over
  rather straightforwardly.  
\item[iv.] In intermediate regimes when we correlate data segments
  separated by a maximum coherence time $\Tmax < \Tobs$, the
  cross-correlation search is similar to a hierarchical search in
  which we combine segments of demodulated data. Though, as discussed
  in Sec.~\ref{sec:params}, there are differences between the two with
  the cross-correlation search being somewhat more sensitive.
\end{description}
Conceptually, this method thus provides a unified framework for all
the known periodic wave searches, and this might be useful in various
calculations and applications.  Each of the above modes of operation
correspond to tuning the maximum coherence time all the way from 
small values to the total observation time.  The precise value chosen
for a specific application depends on the trade-offs between
computational cost, sensitivity, and robustness against signal
uncertainties. The additional parameter which figures importantly in
this trade-off is the length $\Delta T$ of the short data segments.

There are a number of open issues for future work.  An important
question is to get a detailed understanding of the trade-offs
mentioned above for various types of searches including all sky
searches for isolated GW pulsars, signals from known binary systems or
from interesting areas such as the galactic center etc.  This will
help us better decide how to best use our computational resources and
to maximize our chances of making a detection.  Another important
issue, which feeds into this optimization problem, is to study the
general parameter space metric.  To date we only have a proper
understanding of the coherent metric, i.e.  case (i) above.  For the
other cases, we have estimates of the parameter space resolution and
which are often sufficient for many applications, but a full
understanding is still lacking.  In addition, it would be interesting
to compare the estimation of the amplitude parameters
$\{A_+,A_\times\}$ (and $\psi$) obtained from \eqref{eq:77} and
\eqref{eq:78} with the maximum likelihood estimators obtained from the
$\F$-statistic calculation.  In the limit when we consider all
possible correlations, we would expect the two estimates to be very
close to each other.

\acknowledgments

We are grateful to Stefan Ballmer for valuable discussions.  We also
acknowledge all members of the Continuous-Waves working group of the
LIGO Scientific Collaboration for numerous discussions and suggestions
which were crucial for this paper. BK and JTW acknowledge the support
of the Max-Planck-Gesellschaft.  BK acknowledges the University of the
Balearic Islands for hospitality while this work was being carried
out. JTW also acknowledges the support of the German Aerospace Center
(DLR). HM thanks the Council of Scientific and Industrial Research of
India (CSIR) for providing a research scholarship.

\appendix

\section{Including self-correlations and $\mathcal{O}(h_0^2)$
  corrections} 
\label{sec:generalstat}

In this section we relax the two assumptions of only looking at
$\y_{I\!J}$ for $I\neq J$ and $h \ll n$.  We allow self correlations
(which, by themselves, are used in the standard semi-coherent
searches), and we keep terms of $\mathcal{O}(h_0^2)$ but still neglect
$\mathcal{O}(h_0^4)$ terms.  

Let us again start from the general statistic $\rho$ defined in
\eqref{eq:41}, and let us calculate its mean and standard
deviation with the two assumptions relaxed.  In general, we have
$\y_{I\!J}^\conj = \y_{J\!I}$ so that $\y_{I\!I}$ is real and so is the
corresponding weight $u_{I\!I}$; $\y_{I\!I}$ is in fact just the power
in a single SFT bin.  We will denote $\y_{I\!I}$ simply by $\y_I$ and
$u_{I\!I}$ by $u_I$.  

The mean is easy to calculate:
\begin{equation}
  \label{eq:80}
  \langle\y_{I\!J}\rangle := \mu_{I\!J} = \frac{1}{2\Delta T}
  S_n^I\delta_{I\!J} + h_0^2\G_{I\!J}
  \ . 
\end{equation}
Thus, the mean is non-zero in the absence of a signal only for the
self-correlation terms.  In general, $\rho$ will contain
self-correlations, and also correlations of distinct pairs.  However,
we want to be completely general and we do not assume that it contains
\emph{all} the possible pairs.  This is then the expression for the
mean:
\begin{equation}
  \label{eq:81}
  \langle\rho\rangle := \mu = \frac{1}{\Delta T}\sum_{I} u_I S_n^I +
  h_0^2\sum_{\alpha}(u_{\alpha}\G_{\alpha} + u_{\alpha}^\conj\G_{\alpha}^\conj )
  \ .
\end{equation}
It is to be understood that the first sum in this equation only
contains the self-correlations and the second sum contains all the SFT
pairs we have chosen to include, including the self-correlations.

The variance calculation is somewhat more involved.  Before looking at
the variance of $\rho$ itself, let us look at $\langle
\y_{I\!J}\y_{K\!L}\rangle$.  Note that for the pure noise terms:
\begin{equation}
  \label{eq:83}
  \langle
  \tilde{n}_I^\conj\tilde{n}_J\tilde{n}_K^\conj\tilde{n}_L\rangle =
  2\delta_{I(J}\delta_{L)K}\langle|\tilde{n}_I|^2\rangle\langle
  |\tilde{n}_K|^2\rangle
  \ .
\end{equation}
Here, we use the notation that indices within parentheses are
symmetrized over: $X_{(IJ)} = (X_{IJ} + X_{JI})/2$.  This also covers
the $I=J=K=L$ case, so there is no need to consider that separately.

Consider now the signal.  In general, the terms in $\langle
\y_{I\!J}\y_{K\!L}\rangle$ with odd powers of $h$ will vanish because
the noise is assumed to have zero mean. Thus, schematically, we will
have
\begin{equation}
  \label{eq:84}
  \langle \y_{I\!J}\y_{K\!L}\rangle = A + Bh_0^2 + Ch_0^4
  \ .
\end{equation}
Let us ignore the $h_0^4$ terms and focus only on the second order
terms.  The reader can convince herself that we only need to keep the
following terms in $\y_{I\!J}\y_{K\!L}$:
\begin{equation}
  \label{eq:85}
  \tilde{h}_I^\conj\tilde{h}_J\tilde{n}^\conj_K\tilde{n}_L +
  \tilde{h}_K^\conj\tilde{h}_J\tilde{n}^\conj_I\tilde{n}_L +
  \tilde{h}_K^\conj\tilde{h}_L\tilde{n}^\conj_I\tilde{n}_J +
  \tilde{h}_I^\conj\tilde{h}_L\tilde{n}^\conj_K\tilde{n}_J
  \ .
\end{equation}
Putting together \eqref{eq:83} and \eqref{eq:85}, we end up with  
\begin{multline}
  \label{eq:86}
  \langle\y_{I\!J}\y_{K\!L}\rangle = \frac{1}{2(\Delta T)^2}
  \delta_{I(J}\delta_{L)K}{S_n^{(I)}S_n^{(K)}} \\  
  + \frac{h_0^2}{\Delta T}\left[\G_{I(J}\delta_{L)K} S_n^{(K)} +
    \delta_{I(J}\G_{L)K} S_n^{(I)}\right]
  \ .
\end{multline}
We are now ready to look at the variance of $\rho$.  Let us define
$\rho_\alpha = u_\alpha\y_\alpha + u_\alpha^\conj\y_\alpha^\conj$, so
that $\rho = \sum_\alpha\rho_\alpha$.  Then, we have
\begin{equation}
  \label{eq:82}
  \Var\left(\rho\right) = \sum_\alpha
  \Var\left(\rho_\alpha\right) + \sum_{\alpha,\beta\, (\alpha \neq\beta)}
  \Cov\left( \rho_\alpha,\rho_\beta \right)
  \ .
\end{equation}
Let us start with the variances 
\begin{equation}
  \label{eq:88}
  \Var\left(\rho_{I\!J}\right) = \langle \rho_{I\!J}^2 \rangle
  - \mu_{I\!J}^2
  \ .
\end{equation}
For $I\neq J$, $\mu_{I\!J} = \mathcal{O}(h_0^2)$ so that $\mu_{I\!J}^2$
can be ignored.  Thus, in this case we get:
\begin{multline}
  \label{eq:89}
  \sigma_\alpha^2 = \Var\left(\rho_{I\!J}\right)
  = 2|u_{I\!J}|^2 \left\{\frac{S_n^{(I)}S_n^{(J)}}{4\Delta T^2}\right.
  \\\left. + \frac{h_0^2}{2\Delta T}\left(\G_IS_n^{(J)} +
      \G_JS_n^{(I)}\right)\right\}
  \ .
\end{multline}
For the $I=J$ case, we can no longer ignore the $\mu_\alpha$ term.
Keeping terms up to $\mathcal{O}(h_0^2)$ we end up with
\begin{equation}
  \label{eq:91}
  \sigma_{I}^2 =  \Var\left(\rho_{I}\right) = 4u_I^2 \left\{
    \left(\frac{S_n^{(I)}}{2\Delta T}\right)^2 + \frac{h_0^2}{\Delta
      T}\G_IS_n^{(I)}\right\}
  \ .
\end{equation}
Turning now to the covariances, first note that if $I,J,K,L$ are all
distinct, then up to $\mathcal{O}(h_0^4)$ terms,
$\Cov\left(\rho_{I\!J},\rho_{K\!L}\right) = 0$; thus we need
at least one pair of matching indices to get a non-zero result.  Using
\eqref{eq:86} the expressions for all the non-zero cases are the
following ($I\neq J$) ignoring, as always, the $\mathcal{O}(h_0^4)$
terms:
\begin{subequations}
  \begin{align}
    \label{eq:93}
      \langle\y_{I\!I}\y_{I\!J}\rangle &= \frac{h_0^2}{2\Delta
    T}\left(\G_{I\!J}+\G_{J\!I}\right)S_n^{(I)}
  \ ,
  \\
  \langle\y_{I\!I}\y_{J\!I}\rangle &= \frac{h_0^2}{\Delta
    T}\G_{I\!J}S_n^{(I)}
  \ ,
  \\
    \langle\y_{I\!I}\y_{J\!J}\rangle &=
    \frac{S_n^{(I)}S_n^{(J)}}{4\Delta T^2} + \frac{h_0^2}{2\Delta
    T}\left(\G_{I}S_n^{(J)} +\G_{J}S_n^{(I)}\right)
  \ .
  \end{align}
\end{subequations}
It turns out that the only non-zero covariance is 
\begin{equation}
    \label{eq:94}
    \Cov\left(\rho_I,\rho_{I\!J}\right) = \frac{h_0^2}{\Delta
      T} u_IS_n^{(I)}\left( u_{I\!J}\G_{I\!J}^\conj +
      u_{I\!J}^\conj\G_{I\!J}  \right)
    \ .
\end{equation}
We are almost done now. Substituting the results of
\eqref{eq:89}, \eqref{eq:91}, and \eqref{eq:94} in \eqref{eq:82}
we get
\begin{multline}
  \label{eq:95}
  \sigma^2 = 2\sum_{\alpha}
  |u_\alpha|^2\sigma_{(0),\alpha}^2 + \frac{h_0^2}{\Delta T}\Biggl\{ 
  \sum_I4u_I^2\G_IS_n^{(I)} \\ 
  + \sum_{\alpha,I\neq J}|u_{I\!J}|^2\left(\G_IS_n^{(J)} +
    \G_JS_n^{(I)}\right)   \\ 
  + \sum u_I(u_{I\!J}\G_{I\!J}^\conj +
  u_{I\!J}^\conj\G_{I\!J})S_n^{(I)}\Biggr\}
  \ .
\end{multline}
Here we have defined the variances in the absence of a signal:
\begin{subequations}
  \begin{align}
    \label{eq:96}
    \sigma_{(0),I} &= \frac{\left(S_n^{(I)}\right)^2}{2\Delta T^2}
    \ ,
    \\
    \sigma_{(0),I\!J} &= \frac{S_n^{(I)}S_n^{(J)}}{4\Delta T^2}
    \ ,
    \quad
    I\neq J
    \ .
  \end{align}
\end{subequations}
It is convenient to write \eqref{eq:95} in the abbreviated form
\begin{equation}
  \label{eq:97}
  \sigma^2 = \sigma_{(0)}^2 + h_0^2\sigma_{(1)}^2
\end{equation}
where the definitions of $\sigma_0$ and $\sigma_{(1)}$ are obvious
from \eqref{eq:95}.  

We are finally ready to derive the equation for the sensitivity, i.e.
the analogs of \eqref{eq:47} and \eqref{eq:51}. \eqref{eq:45}
for the threshold is unchanged as long as we use $\sigma_{(0)}$
instead of $\sigma$ in that equation\footnote{The mean of $\rho$ is
  now no longer necessarily zero in the absence of a signal (see
  \eqref{eq:81}. But this only leads to an additive correction to
  the threshold $\rho_{\rm th}$, and we assume this correction has
  been made.}. \eqref{eq:46} for the false dismissal rate becomes: 
\begin{equation}
  \label{eq:98}
  \gamma = \frac{1}{2}\erfc
  \left(
    \frac{\rho_{\rm th} - \mu}{\sqrt{2}\sigma}
  \right)
  \ .
\end{equation}
Keeping terms linear in $h_0^2$, we get
\begin{multline}
  \label{eq:99}
  \erfc^{-1}(2\gamma)  = \frac{\rho_{\rm th} -
    \mu}{\sqrt{2}\sigma_{(0)}}\left(1-\frac{h_0^2\sigma_{(1)}^2}{2\sigma_{(0)}^2}\right) \\ 
  = \erfc^{-1}(2\alpha) - \frac{h_0^2}{\sqrt{2}\sigma_{(0)}}
  \sum_\alpha(u_\alpha\G_\alpha + u_\alpha^\conj\G_\alpha^\conj) \\
  - \frac{h_0^2\sigma_{(1)}^2}{2\sigma_{(0)}^2}\erfc^{-1}(2\alpha)
  \ .
\end{multline}
Solving for $h_0$ leads to the generalization of \eqref{eq:47}: 
\begin{equation}
  \label{eq:100}
  h_0^2 = 2\mathcal{S}
  \left(
    \frac{\sqrt{\sum_\alpha |u_\alpha|^2\sigma_{(0),\alpha}^2}}
    {\sum_\alpha(u_\alpha\G_\alpha + u_\alpha^\conj\G_\alpha^\conj)
      + \sigma_{(1)}^2\erfc^{-1}(2\alpha)/\sqrt{2}\sigma_{(0)}}
  \right)  
\end{equation}
Finding the optimal weights is now not as straightforward as before.
However, we note that when $\sigma_{(1)}$ is ignored, then the optimal
weights are again given by \eqref{eq:48} except that now it holds also
for the self-correlations.  In the general case when we do not ignore
$\sigma_{(1)}$, it is simpler to continue using the optimal weights
derived earlier in \eqref{eq:48}, and to substitute it in
\eqref{eq:100} to derive the corresponding sensitivity.

\bibliographystyle{apsrev-titles}

\bibliography{radiometer}

\end{document}